\documentclass[12pt,nohyper]{JHEP3}         % For preprint
\usepackage{amssymb,amsfonts}
\font\mybb=msbm10 at 10pt
\def\bb#1{\hbox{\mybb#1}}

\def\bE {\bb{E}}

\def\bC {\bb{C}}

\newcommand{\be}{\begin{equation}}
\newcommand{\ee}{\end{equation}}
\newcommand{\bea}{\begin{eqnarray}}
\newcommand{\eea}{\end{eqnarray}}

\newcommand{\nn}{\nonumber \\}

\newcommand{\p}[1]{(\ref{#1})}
\newcommand{\lb}[1]{\label{#1}}
\def\6{\partial}
\def\7{\tilde}
\def\8{\widehat}

\def\G11{\Gamma_{11} }
\title{Unitary  Spherical Super--Landau Models}

\author{Andrey Beylin and Thomas L. Curtright\\
         Department of Physics \\
University of Miami, \\
Coral Gables, FL 33124, USA\\
E-mail: beylin@server.physics.miami.edu, curtright@miami.physics.edu \\
}

\author{Evgeny Ivanov\\
    Bogoliubov Laboratory of Theoretical Physics \\
      JINR, 141980 Dubna, Moscow Region, Russia\\ 
 E-mail: eivanov@theor.jinr.ru \\}

 \author{Luca Mezincescu\\
         Department of Physics \\
University of Miami, \\
Coral Gables, FL 33124, USA\\
E-mail: mezincescu@server.physics.miami.edu \\}

\author{Paul K. Townsend\\
Department of Applied Mathematics and Theoretical Physics \\
Centre for Mathematical Sciences, University of Cambridge\\
Wilberforce Road, Cambridge, CB3 0WA, UK\\
E-mail: p.k.townsend@damtp.cam.ac.uk
}

\abstract{A Hilbert space metric is found for the $SU(2|1)$-invariant `superflag' Landau models, parametrized by integer $2N'$
and real number $M$, such that the Hilbert space norm is positive definite. The spectrum of these unitary
super-Landau models is determined. The $M=0$ case yields a unitary Landau model
on the supersphere $SU(2|1)/U(1|1)$ with $U(1)$ charge $2N=2N'+1$.
For the generic unitary superflag model, the manifest $SU(2|1)$ symmetry
is dynamically enhanced to $SU(2|2)$; this is the `spherical' analog
of the hidden worldline supersymmetry found previously in the planar limit.
}
\preprint {DAMTP-2007-120, JINR E2-2008-52, UMTG-6}
\begin{document}
%%%%%%%%%%%%%%%%%%%%%%%%%%%%%%%%%%%%%%%%%%%%%%%%%%%%%%%%%%%%%%%
\section{Introduction}
\setcounter{equation}{0}

A Landau model describes the quantum mechanics of an electrically charged particle confined to
a surface through which passes a constant uniform magnetic flux. In Landau's original paper the surface
was planar, but this may be viewed as the $R\to\infty$ limit of a model in which the surface
is a 2-sphere of radius $R\,$.  In the latter case, the magnetic field can be interpreted
as the field due to a magnetic monopole at the centre of a ball in $\bE^3$ with the 2-sphere
as its surface. Dirac's quantization condition then applies, so that the particle's electric
charge is an integer multiple of a minimal allowed charge. We call this integer $2N\,$,
for reasons to be explained later, and we assume it to be positive. The planar Landau model
is then found by taking the limit in which $R\to\infty$ and $N\to\infty$ with $N/R^2\equiv \kappa$ kept fixed.

In this paper we continue a study of super-Landau models \cite{Ivanov:2003qq,Ivanov:2004yw,Ivanov:2005vh,Curtright:2006aq},
defined either as Landau models on  homogeneous superspaces
that have the 2-sphere as `body', or as planar limits of such models. The spherical super-Landau models
to be considered are those for which the superspace has a transitive action of the supergroup $SU(2|1)\,$,
which allows two possibilities.  The simplest such superspace is the Riemann supersphere:
\be
SU(2|1)/U(1|1) \cong \bC\bb{P}^{(1|1)} \qquad \mbox{ [Supersphere]}\,.
\ee
As for the standard spherical Landau model, there is a family of superspherical Landau models indexed
by a positive integer $2N\,$. In the planar limit one gets the superplane Landau models, indexed
by the real number $\kappa\,$, although  qualitative properties do not depend on this parameter so there
is essentially only one `superplane' model\footnote{Here we should point out that this statement applies
to the supersphere as defined above; an alternative definition yields the alternative `supersphere'
and `superplane' Landau models studied in \cite{Hasebe:2004yp,Hasebe:2004hy,Hasebe:2005cm}
(see \cite{Hasebe:2007gt} for a recent review).}, which we have investigated in detail in two previous papers
\cite{Ivanov:2005vh,Curtright:2006aq}. Excluding this case, we may set to unity the 2-sphere
radius $R\,$,
without loss of generality.

The generic `spherical' super-Landau model with transitive action of $SU(2|1)$ is a `superflag' Landau model,
for which the homogeneous superspace is
\be
 SU(2|1)/[U(1)\times U(1)] \qquad \mbox{ [Superflag]}\,.
\ee Geometrically, the superflag is defined  via the nested
sequence of  superspaces
\be
\bC^{(0|1)} \subset \bC^{(1|1)}
\subset \bC^{(2|1)}\, .
\ee
Each such sequence is a point on the
superflag. The supersphere is then found as the projection in
which one `forgets' the $\bC^{(0|1)}$ superspace.  If instead one
`forgets' the intermediate $\bC^{(1|1)}$ superspace then one gets
the Grassmann odd manifold $SU(2|1)/U(2)$, for which the lowest
Landau level limit was considered  in \cite{Ivanov:2003ax}; we
shall not consider this model in detail here because it is not a
`spherical'  super-Landau model. More about the geometry of flag
supermanifolds may be found in  \cite{Murray:2006pi}.

A class of superflag models, indexed both by the positive integer $2N$ and by another
continuous parameter $M\,$, was constructed in \cite{Ivanov:2004yw}.
Although there is an additional complex anti-commuting coordinate $\xi\,$,
as compared to the supersphere, there is also more freedom in the choice of `kinetic' terms. In fact, there
are now three separate possible $SU(2|1)\,$-invariant kinetic terms that  lead to second time-derivatives
in the equations of motion. One linear combination yields the K\"ahler sigma-model on the  superflag,
but the combination chosen in \cite{Ivanov:2004yw} leads to a degenerate `metric'  for which $\xi$ has
no kinetic term.  The parameter $M$ is the coefficient of a `Wess-Zumino'  term that  involves the
time-derivative of $\xi\,$, so the equation of motion for $\xi$ is generally a first order differential equation,
but it  becomes  algebraic for $M=0$. Provided that the classical energy is non-zero, this allows $\xi$
to be eliminated for $M=0\,$, and the resulting action is precisely that of the
superspherical Landau model. In the planar limit, the subtlety involving zero classical energy has no
effect  in the quantum theory and the $M=0$ planar superflag model is equivalent to the superplane
model  \cite{Ivanov:2005vh,Curtright:2006aq}.  Here we show that a similar equivalence holds for
the spherical Landau models but involving a shift of $2N$ by one unit: the $M=0$ superflag model
with charge $2N'= 2N-1$  is equivalent to the superspherical model with charge $2N\,$.

Although spherical Landau models involve non-linearities absent from the planar models, they are
conceptually simpler in the sense that each Landau level carries a finite-dimensional representation
of the isometry group of the surface on which the particle is moving. This includes the $SU(2)$ isometry
group of the 2-sphere  in all the above cases, but the $SU(2)$ representations must combine into
representations of $SU(2|1)$ in the super-Landau models.  It was shown in \cite{Ivanov:2003qq} that
the lowest Landau level (LLL) of the superspherical  model consists of states that span a degenerate
(atypical) `superspin' $N$ representation of $SU(2|1)\,$; this decomposes into the direct sum of
a  spin $(N\!-\!1/2)$ and a spin $N$ representation of $SU(2)\,$; this is one way to see why $2N$
must be an integer.  A satisfactory definition
of the superspherical Landau model beyond the LLL is complicated by the fact that  each higher
level has states of negative norm (ghosts) in the natural superspace metric \cite{Ivanov:2004yw},
so the naturally defined quantum theory is not unitary.  The problem is less severe for the
superflag models with large positive $M$ because the first  $\left[2M\right] +1$  levels are then
ghost free \cite{Ivanov:2004yw} ($\left[ 2M \right]$ is the integer part of  $2M$). However, there
are still ghosts in the higher levels, and in all levels if $M<0\,$.

In a previous paper, it was shown how this difficulty can be overcome  {\it in the planar limit}
by an alternative choice of  Hilbert space norm \cite{Curtright:2006aq}. In the planar limit,
the $SU(2|1)$ symmetry algebra gets contracted to the  superalgebra $ISU(1|1)\,$, and it turns out that
there are two possible  $ISU(1|1)$ invariant Hilbert space norms, each associated to a choice
of `metric' operator $G\,$.  The trivial choice $G=1$ yields the indefinite Hilbert space norm
but there is a second non-trivial possibility, which yields a positive definite norm for $M\le0\,$,
and one can define a unitary theory for $M>0$ by a `dynamically chosen' mixture of the two invariant
norms.  The changed norm leads to a change in the operation of  hermitian conjugation with
the result that the new hermitian conjugates of the `odd' charges of $ISU(1|1)$ are shifted
by odd operators that are `new' symmetries of the model. Remarkably, these are just worldline
supersymmetry charges, so the unitary  `superplane' Landau model (corresponding to the choice $M=0$)
has a hidden worldline supersymmetry  (as found earlier in \cite{Hasebe:2005cm}
for an alternative superplane Landau model that is apparently {\it quantum}
equivalent to our superplane Landau model). The planar superflag model also
has this worldline symmetry for $M<0\,$, but it is spontaneously broken
\cite{Curtright:2006aq}. A superfield formulation,  in which the worldline ${\cal N}=2$
supersymmetry  is manifest, was presented in \cite{Ivanov:2008}.

The main purpose of this paper is to present results of a similar analysis of the spherical
super-Landau models.  In particular we show that there are two possible $SU(2|1)$ invariant norms on
the  Hilbert space of the superflag Landau model,  each associated to a metric operator  $G$.
The `trivial'  choice $G=1$ yields the indefinite Hilbert space norm but the other choice of $G$
yields a positive Hilbert space norm provided that $-2N' -1<M\le 0\,$, with zero norm states at $M=0\,$;
for other values of $M$ one needs a `dynamical' combination of the two norms. We solve the model
in the sense that we determine the spectrum, degeneracies, and $SU(2|1)$ representations at each
level.  We do the same for the supersphere; in particular, we confirm the earlier result
of \cite{Ivanov:2003qq} that the LLL furnishes an irreducible superspin $N$ representation
of $SU(2|1)\,$.  The results agree with the $M=0$  superflag after taking into account  zero-norm states
of the latter and  the shift of $2N$ mentioned above, and this establishes the equivalence of these
two models.  For the cases in which $-2N' -1 <M\le 0$ we also investigate the nature of
the `hidden' symmetries that are revealed by the process described above for the planar models.
For the supersphere, i.e. $M=0\,$,  we again find additional `supersymmetries' but they
do not form a closed algebra  with the Hamiltonian, except in the planar limit;
it appears likely that closure requires an infinite set  of `new' charges. Thus,
the supersphere Landau model does not have a conventional worldline supersymmetry, in contrast to the superplane model.
 The situation for the  $-2N' -1 <M\le0$  superflag models is rather different, and surprising.
 We show that the manifest $SU(2|1)$ symmetry of these models is enhanced to  $SU(2|2)$,
 with the central charge being a linear function of the Landau level number.

\subsection{Organization}

We will start by formulating the classical superspherical Landau
model. The quantum Hamiltonian is not obviously factorizable, as
it is for the `bosonic' model, but  we nevertheless find an
infinite set  of eigenstates using covariance
arguments\footnote{Our method, which follows the spirit of
\cite{Ferapontov:2000pv} and \cite{Mezincescu:2001bq},  can also
be applied to other models and we reproduce  in an appendix the
results of  Karabali and Nair \cite{Karabali:2002im} for the
$\bC\bb{P}^n$ models, with the advantage that the eigenvectors are found
explicitly without using knowledge of the Wigner  functions for
$SU(n+1)$.}. The next step is to compute the norm of the
eigenvectors. It turns out that this norm can be expressed through
two analytic superfieds  and that its component form is identical
to a particular case of a norm considered for the superflag Landau
model in  \cite{Ivanov:2004yw}.  In this way  it is recognized
that the redefinition of the norm needed for a unitary
superspherical Landau model  is a particular case of the
redefinition needed for the superflag Landau model.

We then turn to the superflag model, reviewing results of
\cite{Ivanov:2004yw}. There is an additional anti-commuting
variable in comparison to the superspherical model, and this leads
to `extended' superfields upon quantization. For all Landau
levels, the eigenvectors are expressed in terms of the components
of a single extended analytic superfield.  We determine the action
of the $SU(2|1)$  charges  in this analytic subspace, and
diagonalize the natural superspace metric  within it. This allows
us to construct the metric operator that ensures a positive
definite norm. As explained above, this can lead to the appearance
of new `hidden' symmetries, and we show that  the
manifest $SU(2|1)$ symmetry is enhanced to $SU\left(2|2\right)$
when $-2N' -1 < M < 0\, $.

Finally, we discuss the relationships between the supersphere and superflag
models using the geometric language of  non-linear realizations
and covariant derivatives on the corresponding supercosets \cite{Ivanov:2004yw}.
In particular, we show that the superflag Hamiltonian of ref.
\cite{Ivanov:2004yw} and that of the supersphere considered here
are two particular cases of  a more general second-order covariant differential operator
defined on the full superflag manifold $SU(2|1)/[U(1)\times U(1)]\,$,
and that  each is recovered after imposing appropriate covariant
conditions on the superfield wave functions. The covariant approach makes explicit
the quantum equivalence of the $2N'$ superflag model  at $M=0$ with
the $2N$ supersphere model when $2N'=2N-1\,$.

The main conclusions of this paper, taken together with our earlier work on
super-Landau models, is summarized in the final section.

%%%%%%%%%%%%%%%%%%%%%%%%%%%%%%%%%%%
\section{The superspherical Landau model}
\setcounter{equation}{0}

We begin with a presentation of some facts about $SU(2|1)$ and the
supersphere. We then construct the classical superspherical Landau
model, solve the quantum model by determining the energy
eigenstates and their eigenvalues. We conclude this section with a
discussion of the Hilbert space norm, noting the problem of
ghosts, which will be resolved following our results for the more
general superflag model.

\subsection{SU(2$|$1)}

The Lie superalgebra  $su(2|1)$ is spanned by even charges
$(F,J_3, J_\pm)$, satisfying the commutation relations of $U(2)$,
and a $U(2)$ doublet of odd charges $(\Pi, Q)$; we write the
complex conjugate charges as  $(\Pi^\dagger,Q^\dagger)$ since we
plan to realize this algebra in terms of operators for which
$(\Pi^\dagger,Q^\dagger)$ are the hermitian conjugates of $(\Pi,
Q)$. The non-zero commutators of the even charges are
\be
\left[J_+,J_-\right] = 2J_3\, , \qquad \left [J_3,J_\pm\right] =
\pm J_\pm\, .
\ee
The non-zero commutators of the odd generators
with the even generators are
\bea
\left[J_+,\Pi\right] &=& iQ\, , \qquad \left[J_-,Q\right]= -i\Pi\, , \nonumber\\
\left[J_3,\Pi\right]  &=& -\frac{1}{2}\Pi\, , \qquad \left[J_3, Q\right]
= \frac{1}{2}Q\,,\nonumber\\
\left[F,\Pi\right]  &=& -\frac{1}{2}\Pi \, , \qquad \left[F, Q\right] = -\frac{1}{2}Q
\eea
and
\bea
\left[J_-,\Pi^\dagger \right] &=& iQ^\dagger\, , \qquad
\left[J_+,Q^\dagger \right]= -i\Pi^\dagger\, , \nonumber\\
\left[J_3,\Pi^\dagger\right]  &=& \frac{1}{2}\Pi^\dagger\, , \qquad
\left[J_3, Q^\dagger\right] = -\frac{1}{2}Q^\dagger\,,\nonumber\\
\left[F,\Pi^\dagger \right]  &=& \frac{1}{2}\Pi^\dagger \, ,
\qquad \left[F, Q^\dagger\right] = \frac{1}{2}Q^\dagger\, , \eea
which show that $(\Pi,Q)$ and $(\Pi^\dagger,Q^\dagger)$ are
$SU(2)$ doublets of  charge $-\frac{1}{2}$ and $\frac{1}{2}$, respectively. Finally,
the non-zero anti-commutators of the odd charges are
\bea
\left\{\Pi,\Pi^\dagger\right\} &=& -J_3 + F \, , \qquad
\left\{Q,Q^\dagger\right\} = J_3 +F\,,  \nonumber\\
\left\{\Pi, Q^\dagger\right\} &=& iJ_- \, , \qquad
\left\{\Pi^\dagger,Q\right\} = -iJ_+\, .
\eea

As $su(2|1)$ is a rank two superalgebra, it  has a quadratic and a  cubic Casimir. The quadratic
Casimir  is
\be
C_2 =\frac{1}{2}\left\{J_+,J_-\right\} + J_3^2
- F^2 -\frac{1}{2}\left[\Pi, \Pi^\dagger \right] -\frac{1}{2}\left[Q, Q^\dagger\right]\, ,
 \lb{Cas2}
\ee
The cubic Casimir operator is
\bea C_3&=& {i\over 2}J_+\left [
Q^\dagger, \Pi\right ] - {i\over 2}\left [ \Pi^\dagger, Q\right
]J_- +{1\over 2}J_3\left( \left [Q,  Q^\dagger \right ]  - \left [
\Pi , \Pi^\dagger \right ] \right) \nn &-&{1\over 2}F\left( \left
[ \Pi , \Pi^\dagger \right ] + \left [Q,  Q^\dagger \right ]
\right)+2C_2 F - \Pi^\dagger \Pi - Q Q^\dagger. \lb{Cas3}
\eea

\subsection{The supersphere}\label{subsec:supersphere}

The Riemann supersphere $\bC\bb{P}^{(1|1)}\cong SU(2|1)/U(1|1)$ is a complex supermanifold with
complex coordinates
\be
Z^A =\left(Z^0,Z^1\right) = \left(z, \zeta\right),  \qquad
\bar Z^{\bar B} = \left(\bar Z^0,\bar Z^1\right) = \left(\bar z, \bar\zeta\right),
\ee
where $z$ is a complex coordinate of the Riemann sphere, with complex conjugate $\bar z$,  and
$\zeta$ is its anti-commuting partner, with complex conjugate $\bar\zeta$.
The $SU(2|1)$ transformations of these complex coordinates are analytic and are generated by the
following differential operators
\bea\label{ssgens}
F &=& \frac{1}{2}\zeta\partial_\zeta\, , \qquad
J_3 = z\partial_z + \frac{1}{2}\zeta\partial_\zeta\,,   \nonumber\\
J_- &=&-i \partial_z\, , \qquad J_+ = -i\left(z^2\partial_z + z\zeta \partial_\zeta \right), \nonumber\\
\Pi &=& \partial_\zeta\, , \qquad \Pi^\dagger = -\zeta z\partial_z\,, \nonumber\\
Q &=& z\partial_\zeta \, , \qquad Q^\dagger = \zeta \partial_z\, .
\eea
The notation suggests that $(\Pi^\dagger,Q^\dagger)$ may be interpreted as hermitian conjugates
of $(\Pi,Q)$, and this is a correct interpretation in the context of the Hilbert space norm for the
superspherical Landau model that we will discuss below.

The infinitesimal $SU(2|1)$ transformations of the coordinates  are found from
\be
\delta Z^A = i\left[ \lambda J_3 + \mu F + \varepsilon J_- + \bar\varepsilon J_+
-i \epsilon^1\Pi -i\bar\epsilon_1\Pi^\dagger  +i \epsilon^2Q +i \bar\epsilon_2Q^\dagger \, , Z^A\right],
\ee
where $\lambda$ and $\mu$ are real,  $\varepsilon$ is complex with complex conjugate
$\bar\varepsilon$, and $(\epsilon^1,\epsilon^2)$ are complex anti-commuting parameters
with complex conjugates $(\bar\epsilon_1,\bar\epsilon_2)$. One finds that
\bea\label{superstr}
\delta  z &=& i\lambda z +  \varepsilon + {\bar \varepsilon}  z^2 - \left(\bar\epsilon_2 +
z\bar\epsilon_1\right)\zeta \, ,\nn
\delta \zeta &=&  \frac{i}{2}\left(\lambda+ \mu\right)\zeta +
\epsilon^1- {\epsilon}^2 z + \bar\varepsilon z \, \zeta  \,  .
\eea
The complex conjugate expressions give the infinitesimal $SU(2|1)$ transformations of $(\bar z,\bar\zeta)$.

The Riemann supersphere is not only a complex supermanifold but
also a K\"ahler supermanifold, with K\"ahler 2-form
\be
{\cal F}
=2 i \, dZ^A \wedge d\bar Z{}^{\bar B}\,  \partial_{\bar
B}\partial_A\,  {\cal K}\, , \ee where \be {\cal K} = \log
\left(1+ z\bar z + \zeta\bar\zeta\right)
\ee
is the K\"ahler
potential, which  is real because the usual convention for complex
conjugation of  products of anti-commuting variables implies  that
$(\partial_\zeta)^*=-\partial_{\bar\zeta}$, and hence that
\be\label{reality}
\left(\partial_{\bar B} \partial_{A} {\cal
K}\right)^* = \left(-1\right)^{a+b} \left(\partial_{\bar B}
\partial_{A} {\cal K}\right).
\ee
Here $a$ is the Grassmann parity associated with the $A$ or $\bar
A$ index; i.e.  $a=0$ for $A=0$ and $\bar A=0$, and $a=1$ for
$A=1$ and $\bar A=1$ (to avoid ambiguities with this simplified
notation, one must arrange for all barred indices to have letters
that differ from those of unbarred indices, but this restriction
is easily accommodated).

The K\"ahler 2-form may be written locally as ${\cal F} =d{\cal A}$, where
\be\label{u1conn}
{\cal A} =-i \left(dZ^A \partial_A - d\bar Z^{\bar B}\partial_{\bar B} \right) {\cal K} \equiv
dZ^A {\cal A}_A + d\bar Z^{\bar B} {\cal A}_{\bar B}
\ee
is the K\"ahler connection. The K\"ahler connection transforms like a $U(1)$ gauge potential under a
K\"ahler gauge transformation ${\cal K} \to {\cal K} + f + \bar f$ for
any analytic function $f$ with complex conjugate $\bar f$, so ${\cal F}$ is K\"ahler gauge invariant.
This implies that it is also $SU(2|1)$ invariant because the $SU(2|1)$ transformation of the K\"ahler potential is
\be
\delta {\cal K} = {\bar\varepsilon} z + \varepsilon {\bar  z} +
\epsilon^1 \bar\zeta  - {\bar \epsilon}_1 \zeta \, ,
\ee
which is a K\"ahler gauge transformation.

The K\"ahler metric of the Riemann supersphere is
\be
d Z^{A} d \bar Z^{\bar B} \,  g_{\bar B A} = d Z^{A} d \bar Z^{\bar B}\,
\partial_{\bar B} \partial_{A} {\cal K} \,.
\ee
It is manifestly K\"ahler gauge invariant, and hence $SU(2|1)$ invariant.  Before proceeding we record, for future use,
the components of the metric and inverse metric.
The metric components are
\bea
g_{\bar z z} &=& \frac{1+ \zeta\bar\zeta}{\left(1+ z\bar z + \zeta\bar\zeta\right)^2}\,,  \qquad
g_{\bar z \zeta} = -\frac{z\bar\zeta}{\left(1+ z\bar z\right)^2}\,,\nn
g_{\bar\zeta z} &=& \frac{\bar z \zeta}{\left(1+z\bar z\right)^2}\,,
\qquad \qquad \quad
g_{\bar\zeta \zeta}= \frac{1}{1+ z\bar z}\, .\lb{Metr}
\eea
The inverse metric components are
\bea
g^{z\bar z} &=& \left(1+z\bar z\right)\left(1+ z\bar z+\zeta\bar\zeta\right), \qquad
g^{z\bar\zeta} = \left(1+ z\bar z\right) z\bar\zeta\,, \nn
g^{\zeta\bar z} &=& -\left(1+ z\bar z\right)\bar z\zeta\,, \qquad \qquad \qquad \quad
g^{\zeta\bar\zeta} = 1+ z\bar z\left(1-\zeta\bar\zeta\right).\lb{Inverse}
\eea

The metric $g_{\bar B A}$ and its inverse $g^{A\bar B}$ are related by the conditions
\be
g^{A\bar B} g_{\bar B C} = \delta^A{}_C \, , \qquad
g_{\bar B C} \, g^{C\bar A} = \delta_{\bar B}{}^{\bar A}\,. \lb{gg-1cond}
\ee

\subsection{The model}

The classical  Lagrangian of the superspherical Landau model is
\be\label{Lag1}
L = \dot Z^A \dot{\bar Z}{}^{\bar B} g_{\bar B A} +N \left(\dot Z^A {\cal A}_A +
\dot {\bar Z}{}^{\bar B}{\cal A}_{\bar B} \right),
\ee
where the overdot indicates differentiation with respect to an independent variable,
which we interpret as time. Observe that $L$ is real as a consequence of (\ref{reality}).
The $SU(2|1)$ variation of this Lagrangian is a total time derivative, for any real number $N$.
As mentioned in the Introduction,  the quantum theory requires $2N$ to be an integer,
which can be interpreted as the particle's electric charge.

We will proceed directly to  the Hamiltonian form of the Lagrangian,
\be\label{sslag}
L = \dot Z^A P_A + \dot {\bar Z}{}^{\bar B} P_{\bar B} - \left(P_A -N{\cal A}_A \right)
g^{A\bar B} \left(P_{\bar B}- N{\cal A}_{\bar B}\right),
\ee
where the inverse metric is defined in \p{Inverse}, \p{gg-1cond}
and the  conjugate momenta are
\be
P_A = \left(p_z, -i\pi_\zeta\right), \qquad
P_{\bar B} = \left(p_{\bar z} , -i \pi_{\bar \zeta} \right).
\ee
Here, $p_{\bar z}$ is the complex conjugate of $p_z$ and $\pi_{\zeta}$ is the complex conjugate
of $\pi_{\bar\zeta}$; the factors of $-i$ are needed for this to be the case as a consequence
of the rule for complex conjugation of products of anti-commuting variables, and this has the consequence that
\be
\left(P_A\right)^* = \left(-1\right)^a  P_{\bar A}\, .
\ee
Since the inverse metric behaves in the same way as the metric under complex conjugation, one sees that
the new Lagrangian, in Hamiltonian form, is real, and one may verify that elimination of the momenta returns
us to the Lagrangian (\ref{Lag1}). We may now read off the classical Hamiltonian, which we rewrite as
\be
H_{class} = \left(-1\right)^{a(a+b)} g^{A\bar B} \left(P_A -N{\cal A}_A \right)
\left(P_{\bar B}- N{\cal A}_{\bar B}\right).
\ee

To  quantize, we make the replacements
\begin{equation}
p_z\rightarrow -i \partial_{z}\, ,\qquad
p_{\bar z}\rightarrow -i\partial_{\bar z},\qquad
\pi_\zeta\rightarrow\partial_{\zeta}\, ,\qquad
 \pi_{\bar \zeta}\rightarrow\partial_{\bar{\zeta}}\,, \label{replaceferm}
\end{equation}
which imply
\be\label{quantP}
P_A \to -i\partial_A\, , \qquad P_{\bar B} \to -i \partial_{\bar B}\, .
\ee
This yields the quantum Hamiltonian
\be
H= -  \left(-1\right)^{a(a+b)} g^{A\bar B}\nabla^{\left(N\right)}_A \nabla^{\left(N\right)}_{\bar B} \, ,\lb{SSham}
\ee
where
\be
\nabla^{\left(N\right)}_A =\partial_A -N\left(\partial_A{\cal K}\right),\qquad
\nabla^{\left(N\right)}_{\bar B} = \partial_{\bar B} +N\left(\partial_{\bar B}{\cal K}\right).\lb{nablas}
\ee
These covariant derivatives have the super-commutator
\be\label{identity1}
\nabla_{\bar B}^{(N)}\nabla_A^{(\tilde N)} - \left(-1\right)^{ab} \nabla_A^{(\tilde N)}\nabla_{\bar B}^{(N)}
= -\left(N+\tilde N\right)g_{\bar B A}\, ,
\ee
with all other super-commutators equal to zero. For further use, we present here the explicit
expressions for $\nabla_A^{(N)}, \nabla^{(N)}_{\bar B}$:
\bea
&& \nabla_z^{(N)} = \partial_z - N \,\frac{\bar z}{1 + z\bar z + \zeta\bar{\zeta}}\,,
\quad \nabla_{\bar z}^{(N)} = \partial_{\bar z} + N \,\frac{z}{1 + z\bar z + \zeta\bar{\zeta}}\,, \nn
&& \nabla_\zeta^{(N)} = \partial_\zeta - N \,\frac{\bar{\zeta}}{1 + z\bar z + \zeta\bar{\zeta}}\,,
\quad \nabla_{\bar{\zeta}}^{(N)} = \partial_{\bar{\zeta}} - N \,\frac{\zeta}{1 + z\bar z + \zeta\bar{\zeta}}\,. \lb{explNab}
\eea

The $SU(2|1)$ invariance of the model can be made manifest by writing the Hamiltonian operator in
terms of the Casimir operators. One finds that
\be
H=  C_2\, .\lb{HC2}
\ee

\subsection{The spectrum}
\label{sect:spectrum}

The energy levels of the Landau model on the sphere may be found
exactly, e.g. using a factorization method. Although it is not
clear to us how to apply this method to  supersphere, the
`supersymmetrization'  will obviously expand the $SU(2)$
representation content at each level to some representation of
$SU(2|1)$. Moreover, the lowest Landau level (LLL) is known from
earlier work \cite{Ivanov:2003qq}; in the present context, in
which we have chosen an operator ordering such that the ground
state energy is zero, the LLL wave functions are components of a
superfield  $\Psi_0^{(N)}$, satisfying the analyticity constraint
\be
\nabla_{\bar B}^{(N)} \Psi_0^{(N)} =0\, ,\lb{LLLSS}
\ee
and they carry
an irreducible superspin $N$ representation of $SU(2|1)$ that
decomposes into the reducible  $(N-1/2)\oplus N$ representation of
$SU(2)$.  More generally, the  energy eigenvalues are
\be\label{casfunction}
E_\ell = C_2(\ell) = \ell\left(\ell + 2N \right)
\ee
for non-negative integer $\ell$, and the states in
the $\ell$th Landau level, for $\ell>0$, have superfield wave functions
of the form
\be
\Psi_{\ell}^{(N)} = \nabla_{A_1}^{(N+1)}  \cdots
\nabla_{A_\ell}^{(N+2\ell-1)} \Phi^{A_\ell \dots A_1}\, ,\lb{1}
\ee
where the superfield $\Phi^{A_\ell \cdots  A_1}$ is totally
graded symmetric in its $\ell$ indices and satisfies the analyticity
condition
\be \nabla_{\bar B} ^{(N)}  \Phi^{A_\ell \dots A_1} =0\,.\lb{2}
\ee
The graded symmetry means that $\Phi$ has only two
independent components, which we may take to be
\be
 \Phi^{z \dots z} \equiv  \Phi_\ell^{(+)}\, , \qquad  \Phi^{z  \dots \zeta} =
 \Phi_\ell^{(-)}\, .\lb{3}
\ee
It follows that
\be
\Psi_{\ell}^{(N)} = \Psi_{(+)\ell}^{(N)} +
\Psi_{(-)\ell}^{(N)}\,, \lb{Psi}
\ee
where the two independent superfields
$\Psi_{(\pm)\ell}^{(N)}$ are given by
\be
\Psi_{(+)\ell}^{(N)} =
\nabla_z^{(N+1)}  \cdots \nabla_z^{(N+2\ell-1)}\,  \Phi_\ell^{(+)}
\lb{4} \ee and \be \Psi_{(-)\ell}^{(N)} = \left[\sum_{p=1}^\ell
\nabla_z^{(N+1)} \dots \nabla_\zeta^{(N+2p-1)} \dots
\nabla_z^{(N+2\ell-1)} \right]\Phi_\ell^{(-)}\, .\lb{5}
\ee
The LLL
is exceptional in that only the $(+)$ component is defined, and this
is the ground state wave function that we called $\Psi_0^{(N)}$. In
general, both of the $\Psi_{(\pm)}$ components  will carry an
irreducible representation of $SU(2|1)$, so only the LLL has a
representation carried by a single analytic superfield. We
arrived at this result  using insights gained from earlier studies
of the planar limit, and by analogy with the $\bC\bb{P}^2$ Landau model,
which we discuss in an appendix. Here we shall verify the result for
the first two levels; this is the beginning of a general inductive
argument, which we shall not present but which should become clear.

At $\ell=1$ we have the superfield wave function
\be
\Psi_1^{(N)} = \nabla_C^{(N+1)}\Phi^C \, .
\ee
After acting with $H$ on this wave function, we move the $\nabla_{\bar B}^{(N)}$ derivative
to the right, where it annihilates $\Phi^C$, but we pick up a super-commutator term, which we simplify
using (\ref{identity1}). The result is
\be
H\Psi_1^{(N)} = (2N+1) g^{a\bar B} \nabla_A^{(N)} g_{\bar B C} \Phi^C\, .
\ee
Now we use  the identity
 \be\label{summed}
  \left(-1\right)^{a(a+b)} g^{A{\bar B}}\nabla^{\left(N\right)}_Ag_{{\bar B}C}= \nabla^{\left(N+1\right)}_C\, ,
 \ee
which itself is a consequence of the identity
 \be
 \left(-1\right)^{a(a+b)}  g^{A{\bar B}}\left({\partial_A}\, g_{{\bar B} C}\right)=-\partial_C \ {\cal K}\, .
\ee
The result is that $\Psi_1^{(N)}$ is an eigenfunction of $H$ with energy eigenvalue $(2N+1)$.

At $\ell=2$ we have the superfield wave function
\be
\Psi_2^{(N)} = \nabla^{\left(N+1\right)}_D \nabla^{\left(N+3\right)}_C \Phi^{CD}\, .
\ee
After acting with $H$ on this superfield we again move $\nabla_{\bar B}^{(N)}$ to the right,
where it annihilates the chiral superfield $\Phi$, but we now pick up two super-commutator
terms. Simplifying these with (\ref{identity1}),  we find that
\bea\label{Hphi}
H \Psi^{(N)}_{(+)2}  &=&  \left(-1\right)^{a(a+b)} \left(2N+1\right)  g^{A\bar B} \nabla_A^{(N)}
g_{\bar B D} \nabla_C^{(N+3)} \, \Phi^{CD} \nn
&&+\, \left(-1\right)^{a(a+b)+bd}\left(2N+3\right)g^{A\bar B} \nabla_A^{(N)}
\nabla_D^{(N+1)}g_{\bar B C} \, \Phi^{CD}\, .
\eea
Now we use the identity
\be\label{identitynabg1}
 \left(-1\right)^{bc} \nabla_{(C}^{(N+1)} g_{\bar B D)} \equiv   g_{\bar B (C}\nabla_{D)}^{(N+3)} \, ,
\ee
where the brackets indicate graded symmetrization in the unbarred  indices, to rewrite (\ref{Hphi}) as
\be
H \Psi^{(N)}_{(+)2} =  \left(-1\right)^{a(a+b)} \left(4N+4\right) g^{A\bar B}
\nabla_A^{(N)} g_{\bar B D} \nabla_C^{(N+3)} \,  \Phi^{CD}\, .
\ee
Then, using (\ref{summed}), we confirm that $\Psi_2^{(N)}$ is an eigenfunction of $H$ with
energy eigenvalue $(4N+4)$.  No new identities are needed to repeat these steps at higher levels,
and the result for the $\ell$th level  may be obtained by induction. In Section 5 we shall reproduce
the same spectrum in an equivalent manifestly
$SU(2|1)$ covariant approach based on the standard non-linear realizations definition
of covariant derivatives on (super)cosets \cite{Ivanov:2004yw}.

We conclude this Section  with a comment.  Observe that in all the above formulas
the derivatives $\nabla_A^{(\tilde N)}, \nabla^{(\tilde N)}_{\bar A}\;(\tilde N'= N, N+1, \ldots)$ are defined
by eqs. \p{nablas}, \p{explNab}: their variations under the odd part of the $SU(2|1)$
coordinate transformations \p{superstr} (and the conjugate ones)\footnote{It suffices
to consider only the transformations with odd parameters, as those with the even parameters
are contained in the closure of those with the odd parameters.} are
\bea
\delta \nabla^{(\tilde N)}_z &=& (\bar\epsilon_1\zeta)\nabla^{(\tilde N)}_z + \epsilon^2\nabla^{(\tilde N)}_\zeta\,, \nn
\delta \nabla^{(\tilde N)}_{\bar z} &=& -(\epsilon^1\bar\zeta)\nabla^{(\tilde N)}_{\bar z} + \bar\epsilon_2\nabla^{(\tilde N)}_{\bar\zeta}\,, \nn
\delta \nabla^{(\tilde N)}_\zeta &=& -(\bar\epsilon_2 + z\bar\epsilon_1)\nabla^{(\tilde N)}_z - \tilde N\bar\epsilon_1\,, \nn
\delta \nabla^{(\tilde N)}_{\bar\zeta} &=& (\epsilon^2 + \bar z\epsilon^1)\nabla^{(\tilde N)}_{\bar z} - 
\tilde N\epsilon^1\, . \lb{Nabtrans}
\eea
Now  observe that the variations of $\nabla^{(\tilde N)}_\zeta$ and  $\nabla^{(\tilde N)}_{\bar\zeta}$ contain pieces $\sim \tilde N\,$.
For the chirality conditions \p{LLLSS}, \p{2} to be covariant, we are led to ascribe similar terms
to the transformations of the wave functions $\Psi^{(N)}_0$ and $\Phi^{(\pm)}_{\ell}$:
\bea
&& \delta \Psi^{(N)}_0 = -N\,(\epsilon^1\bar\zeta + \bar\epsilon_1\zeta)\,\Psi^{(N)}_0\,, \lb{tranLLL} \nn
&& \delta \Phi^{(+)}_{\ell} = -N\,(\epsilon^1\bar\zeta + \bar\epsilon_1\zeta)\,\Phi^{(+)}_{\ell}
 -\ell(\bar\epsilon_1\zeta)\,\Phi^{(+)}_{\ell} +\ell (\bar\epsilon_2 + z\bar\epsilon_1)\,\Phi^{(-)}_{\ell}\,, \nn
&& \delta \Phi^{(-)}_{\ell} =-N\,(\epsilon^1\bar\zeta + \bar\epsilon_1\zeta)\,
\Phi^{(-)}_{\ell}-(\ell-1) (\bar\epsilon_1\zeta)\,\Phi^{(-)}_{\ell}
 +\epsilon^2\, \Phi^{(+)}_{\ell}. \lb{+-transf}
\eea
As expected, the functions $\Psi_{(\pm)\ell}^{(N)}$ defined in \p{4} and \p{5} are not separately covariant
under the transformations \p{Nabtrans} and \p{+-transf}, while the function $\Psi_{\ell}^{(N)}$ defined in \p{Psi}
has a simple transformation law, the same as that of $\Psi^{(N)}_0$:
\be
\delta \Psi_{\ell}^{(N)} = -N\,(\epsilon^1\bar\zeta + \bar\epsilon_1\zeta)\,\Psi_{\ell}^{(N)}\,.\lb{tranLL}
\ee
The weight factor $\sim N$ in \p{tranLLL} - \p{tranLL} is imaginary, so $\vert\Psi^{(N)}_\ell\vert^2
 = \left(\Psi^{(N)}_\ell\right)^*\Psi^{(N)}_\ell$ is a genuine scalar.

\subsection{Hilbert space norm}

The Hilbert space has a natural $SU(2|1)$-invariant norm, defined as the superspace integral \cite{Ivanov:2003qq}
\be\label{ssnorm}
||\Psi||^2 = \int \! d\mu_0  \, e^{-{\cal K}} \,  \Psi^* \Psi \, ,
\ee
where
\be
d\mu_0 = dz d\bar z\,  \partial_\zeta\partial_{\bar\zeta}\, .\lb{meas1}
\ee
For the ground state this norm reproduces the results in \cite{Ivanov:2003qq}. For the first excited
state we may simplify the norm by means of the  integration by parts identity
\be
\int d\mu_0 \, e^{-{\cal K}} \left(\nabla_A^{(N)}\Phi^A\right)^* \Theta \equiv
-\left(-1\right)^{a} \int d\mu_0 \,
e^{-{\cal K}} \left(\Phi^A\right)^* \left(\nabla_{\bar A}^{(N-1)}\Theta\right),
\ee
valid for arbitrary superfield $\Theta$.  Using also the super-commutator identity
(\ref{identity1}) and the chirality condition on $\Phi^C$, we find that
\be
||\Psi^{\left(N\right)}_1||^2 = \left(-1\right)^a \left(2N + 1\right) \int d\mu_0\,  e^{-{\cal K}}
\,   \left(\Phi^B\right)^* \  g_{{\bar B} A }\Phi^{A}.
\ee
Similar steps may be used to simplify the norm of $\Psi_\ell^{(N)}$  for $\ell>1$,
but one now needs the identity, analogous to  (\ref{identitynabg1}),
\be \label{identitynabg2}
\left(-1\right)^{bc}\nabla_{(\bar A}^{(N+2)} g_{\bar B )C}  \equiv g_{(\bar A C} \nabla_{\bar B )}^{(N)}\, ,
\ee
where the brackets again indicate graded symmetrization, but now in the barred indices.
The final result is
\be
||\Psi^{\left(N\right)}_\ell||^2 = \sigma_\ell \frac{(2N+2\ell -1)! \ell!}{2N+\ell -1)!}  \int d\mu_0\,  e^{-{\cal K}} \,
\left(\Phi^{B_1\dots B_\ell}\right)^*
g_{\bar B_1 A_1} \cdots g_{\bar B_\ell A_\ell}\,  \Phi^{A_\ell \dots A_1} \, ,
\ee
where
\be
\sigma_\ell= \left(-1\right)^{\sum_i^\ell b_i + \sum_i^{\ell-1} a_ib_{i+1}}\, .
\ee
In terms of the two independent chiral superfields $\Phi^{(\pm)}_\ell$, we have\footnote{Although
the $\ell=0,1$ cases are special, and need to be considered separately, this result for $\ell\ge2$
is also correct for $\ell=0,1$. In particular,  all terms involving $\Phi^{(-)}$ are absent for $\ell=0$, as expected.}
\bea\label{normphis}
||\Psi^{\left(N\right)}_\ell||^2 &=& {\left( 2N + 2\ell -1\right)!\ell! \over \left( 2N + \ell -1\right)!}
\int \! d\mu_0\,  e^{-{\cal K}}\bigg\{ \left(\Phi_\ell^{(+)} \right)^* \left(g_{{\bar z} z}\right)^{\ell}\Phi_\ell^{(+)} \\
&&+\ \ell \left(\Phi_\ell^{(+)}\right)^* \left(g_{{\bar z} z}\right)^{\ell - 1}g_{{\bar z} \zeta}\, \Phi_\ell^{(-)}
- \ell \left(\Phi_\ell^{(-)}\right)^* \left(g_{{\bar z} z}\right)^{\ell - 1}g_{{\bar \zeta} z}\, \Phi_\ell^{(+)}\nn
&& +\ \left(\Phi_\ell^{(-)}\right)^*
\left[-\ell\left(g_{{\bar z} z}\right)^{\ell - 1}g_{{\bar \zeta} \zeta} +\ell\left(\ell - 1\right)
\left(g_{{\bar z} z}\right)^{\ell - 2}g_{{\bar z} \zeta}
g_{{\bar \zeta} z}\right]\Phi_\ell^{(-)}\bigg\}. \nonumber
\eea

To proceed, we solve the analyticity constraint \p{2} on the $\Phi^{(\pm)}_\ell$ superfields by writing
\be
\Phi_\ell^{(\pm)} =  e^{-N{\cal K}} \varphi^{(\pm)}_\ell\, ,
\ee
where $ \varphi^{(\pm)}_\ell$ are unconstrained analytic superfields, with holomorphic $SU(2|1)$ transformations that follow from (\ref{+-transf}):
\bea
\delta \varphi^{(+)}_\ell &=&  -(2N +\ell)\,(\bar\epsilon_1\zeta)\, \varphi^{(+)}_\ell + \ell \,(\bar\epsilon_2
+ z\bar\epsilon_1)\,\varphi^{(-)}_\ell\,, \nn
\delta \varphi^{(-)}_\ell &=&
-(2N +\ell-1)\,(\bar\epsilon_1\zeta)\, \varphi^{(-)}_\ell
+ \epsilon_2\,\varphi^{(+)}_\ell\,.
\eea
We may expand $ \varphi^{(\pm)}_\ell$ in component fields as follows:
\be
\varphi^{(-)}_\ell = A_\ell +\zeta\psi_\ell \, , \qquad
\varphi^{(+)}_\ell = \chi_\ell +\zeta F_\ell\, .
\ee
If (as the notation suggests) the component functions $(\chi,\psi)$  are assumed to  be Grassmann odd,
and the component functions $(A,F)$ are assumed to be Grassmann even, then $\Psi$ will be Grassmann odd.
With the reverse Grassmann parity assignments to the component functions, $\Psi$ will have even Grassmann parity.
In either of these two cases the `Hilbert' space is actually a supervector space rather than a vector space.
If, instead, all component functions are assumed to be Grassmann even then $\Psi$ will not have
a definite Grassmann parity but the Hilbert space will be a standard Hilbert space. There is
no need here to choose between these alternatives as long as we are careful
not to perform any re-ordering that would require us to specify one of them. Substituting for
$\Phi^{(\pm)}_\ell$ in (\ref{normphis}) and performing the Berezin integration, we arrive at the result
\bea
||\Psi^{\left(N\right)}_\ell||^2 &= &{\left( 2N + 2\ell -1\right)!\ell! \over \left( 2N + \ell -1\right)!}
\int {d z d{\bar z}\over \left( 1 + z{\bar z}\right)^{2\left(N +\ell\right) +1}}
\Bigg[-\ell\left(2N +\ell\right)|A_\ell|^2 -\ell\bar \psi_\ell\psi_\ell \nn
&&- \, \ell\left( \bar \chi_\ell +\bar z\bar \psi_\ell\right)\left(\chi _\ell+ z\psi_\ell\right)
+ \frac{ 2\left(N+\ell\right) + 1}{ 1 + z\bar z}  \bar \chi_\ell \chi_\ell  + |F_\ell|^2\Bigg]\, .
\eea
If in this norm we substitute $2N = 2N' +1$,  we get  the norm found in  \cite{Ivanov:2004yw}
for the $M=0$ superflag Landau model with charge $2N'$. We shall study the general
superflag Landau model in the  following Section, but this result already allows us
to anticipate its equivalence at $M=0$ to the supersphere model, with a shift of the charge by one unit.

The above norm is $SU(2|1)$ invariant, by construction, but not positive definite, so the associated quantum theory
is not unitary. However, there could be an alternative $SU(2|1)$ invariant norm that
{\it is }positive-definite. Indeed there is, but we shall investigate this in the context
of the more general superflag model since we may then specialize to $M=0$ to get
a unitary superspherical Landau model.  Quite apart from the fact that we will then have
the main result in the context of a more general model, another reason for this approach
to the problem is that computations are easier for the superflag model. This is because the additional
anti-commuting variable of the classical theory  becomes an additional superspace coordinate
in the quantum theory, and expansion in this coordinate yields $(\pm)$ pairs of superfields of the type
that we have been considering. This simplification also allows the superflag model to be solved exactly
by a factorization trick.

%%%%%%%%%%%%%%%%%%%%%%%
\section{The superflag Landau model}
\setcounter{equation}{0}

The superflag is the coset superspace $SU(2|1)/[U(1)\times U(1)]$. It is a complex
supermanifold and we may choose
\be
Z^M = (z, \zeta,\xi)\, , \qquad {\bar Z_M }= (\bar z, \bar \zeta, \bar\xi)
\ee
as the complex coordinates, where $(z,\zeta)$ are the complex coordinates used
previously for the supersphere, with $SU(2|1)$ transformations  (\ref{superstr}), and
$\xi$ is a new complex anti-commuting  coordinate
with $SU(2|1)$ transformation
\be
\delta \xi = -\frac{i}{2}\left(\lambda -\mu \right)\xi + \epsilon^2 - \bar\varepsilon \zeta + \left(\bar\epsilon_1\zeta
-\bar\varepsilon z\right)\xi\,.\lb{xiTran}
\ee
The superflag is also a K\"ahler supermanifold, but the K\"ahler metric is {\it not} used in the superflag Landau model,
as constructed in \cite{Ivanov:2004yw}. Instead one uses another $SU(2|1)$-invariant  second-rank tensor field,
a {\it degenerate} one such that there is no `kinetic' term for the new variable
$\xi$. Specifically, the `kinetic' part of the Lagrangian  is constructed from a complex
$SU(2|1)$-invariant  super one-form that induces a worldline one-form with
the coefficient\footnote{This is equivalent to the expression of  \cite{Ivanov:2004yw}, which
is given there in different coordinates.}
\be
\omega^+ =  K_2^{-1}K_1^{-\frac{1}{2}}\left\{ \dot z \left[1- z\xi\bar\zeta - K_2 \xi\bar\xi\right]
- \dot\zeta\left[ z\bar\zeta + K_2\bar\xi\right]\right\},
\ee
where
\be\label{Ks}
K_{1} = 1 + \left(\bar\zeta + \bar z \bar\xi\right)\left(\zeta + z\xi\right) + \bar\xi\xi\, , \qquad
K_2 =  1 +{\bar z}z + \zeta\bar\zeta\, .
\ee
In addition, the model uses the two real $SU(2|1)$-invariant  super 2-forms
\bea
F_1 &=& 2i dZ^M\wedge d\bar Z^{\bar N}\partial_{\bar N}\partial_M \log K_1 = d{\cal B}\nonumber\\
F_2 &=&- 2i dZ^M\wedge d\bar Z^{\bar N}\partial_{\bar N}\partial_M \log K_2 = d{\cal A}\, ,
\eea
where
\bea
{\cal B} &=& i\left(dZ^M\partial_M - d\bar Z^{\bar M}\partial_{\bar M}\right)\log K_1 =
dZ^M{\cal B}_M + d\bar Z^{\bar M}{\cal B}_{\bar M} \, , \nonumber\\
{\cal A} &=&- i\left(dZ^M\partial_M - d\bar Z^{\bar M}\partial_{\bar M}\right)\log K_2
= dZ^M {\cal A}_M + d\bar Z^{\bar M} {\cal A}_{\bar M}\, .
\eea
The $SU(2|1)$-invariance of $F_1$ follows directly from the transformation law
\be
\delta \left(\log K_1\right)= \left(\bar\epsilon_1 \zeta - \epsilon^1\bar\zeta\right)
+ \left(\bar\epsilon_2 + z\bar\epsilon_1\right)\xi
- \left(\epsilon^2 + \bar z\epsilon^1\right)\bar\xi\, .
\ee
The super 2-form $F_2$ is the K\"ahler 2-form of the supersphere model since $\log K_2$
is the K\"ahler potential ${\cal K}$ of the supersphere. Consequently, ${\cal A}$ is the $U(1)$ connection used
in the construction of the supersphere model; in particular, its $\xi$ component is zero,
and the non-zero components are $\xi$-independent.

We now have all the ingredients needed for the generalization from the supersphere Landau model
to the superflag Landau model. The superflag Lagrangian is
\be
L= \left|\omega^+\right|^2 + \left[\dot Z^M \left(N'{\cal A}_M + M{\cal B}_M\right) + c.c.\right], \lb{LagSF}
\ee
where $N'$ and $M$ are two real numbers. In the quantum theory, $M$ remains arbitrary but
$2N'$ must be an integer; we will later see that  the $M=0$ superflag model
is {\it quantum} equivalent to the supersphere model  when $2N =2N'+1\,$, but let us first consider
the relation between the classical Lagrangians of these models. When $M=0$ there are no terms involving time
 derivatives of  $\xi$ in  (\ref{LagSF}),  so the equation of motion of this variable is algebraic.
 By making explicit the $\xi$ dependence in the Lagrangian, one finds that the $\xi$ equation of motion, for $M=0\,$, is
\be\label{algebraic}
\left[\left(1-\bar\zeta\zeta\right)\left|\dot z\right|^2 + \dot z \bar z \dot{\bar\zeta} \zeta
- \dot{\bar z}z\dot\zeta\bar\zeta + \dot\zeta\dot{\bar\zeta} \left(K_2-\bar\zeta\zeta\right)\right] \xi
= -\dot\zeta\left[\dot{\bar z}\left(1-\bar\zeta \zeta\right) + \bar z \dot{\bar\zeta}\zeta\right].
\ee
As long as $\dot z\ne0$, one may use this equation to eliminate $\xi$, in which case the resulting
Lagrangian is equivalent to the Lagrangian for the superspherical Landau model, with $N'=N\,$.
However, when $\dot z=0$, (\ref{algebraic}) is equivalent to
\be
\dot\zeta\dot{\bar\zeta} \left[\left(K_2 -{\bar \zeta}\zeta\right) \xi + \bar z \zeta\right]=0\, ,
\ee
so the solution for $\xi$ is no longer unique but involves terms proportional
to $\dot\zeta$ and $\dot{\bar\zeta}$ with {\it arbitrary} functions as coefficients.
As we will see shortly, this feature is associated to a fermionic gauge invariance
of the $M=0$ superflag model when restricted to configurations with zero energy.

As the last topic of this subsection we note that the holomorphic superspace $(z, \zeta, \xi)$
can be extended to the following complex supermanifold
\be
(z, \zeta, \xi, \widehat{\bar{\xi_{\,}}}), \qquad \widehat{\bar{\xi_{\,}}} = \bar\xi\,K_2 + \bar\zeta\,z\,,\lb{Ext1}
\ee
which is still closed under the action of $SU(2|1)$:
\be
\delta\widehat{\bar{\xi_{\,}}} = \frac{i}{2}\,(\lambda -\mu)\,\widehat{\bar{\xi_{\,}}}
+ \bar\varepsilon z\,\widehat{\bar{\xi_{\,}}}+ \bar\epsilon_2
+ \bar\epsilon_1\,(z + \widehat{\bar{\xi_{\,}}}\,\zeta)\,. \lb{barxiTran}
\ee
This extension of the holomorphic supersphere $(z, \zeta)$ will be exploited in Section 5
where we revisit  the relationships between the supersphere and superflag Landau models.

\subsection{Hamiltonian}

We now turn to a Hamiltonian analysis of the general superflag model.  The model has four primary constraints,
which occur in two complex conjugate pairs. One pair is
\be\label{firstpair}
\varphi_\zeta = {\cal P}_\zeta +i({\bar \xi}K_2 +{\bar \zeta}z){\cal P}_z,\qquad\varphi_{\bar \zeta} = {\cal P}_{\bar \zeta}
 -i({ \xi}K_2 +{ \zeta}{\bar z}){\cal P}_{\bar z}\, ,
 \ee
 where
 \bea
{\cal P}_\zeta  =\pi_\zeta -iN'{\cal A}_\zeta -iM{\cal B}_\zeta ,\qquad  {\cal P}_z = (p_z - N'{\cal A}_z -M {\cal B}_z), \nn
{\cal P}_{\bar \zeta}  = \pi_{\bar \zeta}  -iN'{\cal A}_{\bar \zeta} -iM{\cal B}_{\bar \zeta} ,\qquad  {\cal P}_{\bar z}=
(p_{\bar z} - N'{\cal A}_{\bar z} -M {\cal B}_{\bar z})\, .
 \eea
The other pair is
\be\label{secondpair}
\varphi_\xi = \pi_\xi -iM{\cal B}_\xi, \qquad \varphi_{\bar \xi} = \pi_{\bar \xi} -iM{\cal B}_{\bar \xi}\, .
\ee
The Hamiltonian is
\bea
H_0 = K_2^2K_1^{- 1}\left[1 +\left({\bar \zeta} +{\bar z}{\bar \xi}\right)\zeta\right]\left[1 +{\bar \zeta}\left({ \zeta}
+{ z}{ \xi}\right)\right] {\cal P}_z{\cal P}_{\bar z}\, ,
\eea
where the subscript  is a reminder that we may add  any  function on phase space that  vanishes on the subspace
specified by the primary constraints. A remarkable feature of this Hamiltonian  is that it is independent of $M$.
When we pass to the quantum theory, this means that the energy levels are independent of $M$ but this does not
mean that the parameter $M$ is irrelevant because it can affect the norms of the quantum states.
This effect has a classical  counterpart that we now explain.

A computation shows that  the analytic constraint
functions $\left( \varphi_\zeta, \varphi _\xi \right)$ have
vanishing Poisson brackets among themselves, but that the matrix
of Poisson brackets of these functions with their complex
conjugates is non-zero. In fact,
\be \det \pmatrix{
\{\varphi_\zeta,\varphi_{\bar \zeta}\}_{PB} &
\{\varphi_\zeta,\varphi_{\bar \xi}\}_{PB} \cr
\{\varphi_\xi,\varphi_{\bar \zeta}\}_{PB}  &
\{\varphi_\xi,\varphi_{\bar \xi}\}_{PB} } =
-\left(1+\zeta\bar\zeta\right)\left(1+K_2\,
\bar\xi\xi\right)\left[H_0 -4M\left(N'+M\right)\right].
\ee
It
follows that there is a gauge invariance on the surface in phase
space with energy
\be
H_0 = 4M\left(N' + M\right).
\ee
Indeed, if the determinant of the constraints is weakly zero then the matrix of Dirac brackets
of the constraint functions is degenerate and some constraints must be `first class' ,
in Dirac's terminology, and according to Dirac's formalism there is a gauge invariance
for each first class constraint.  As the constraints are Grassmann odd in our case,
the gauge invariances have Grassmann odd parameters. This generalizes the analogous result of
\cite{Ivanov:2005vh} for the planar superflag model. From the analysis of the planar limit,
we expect that this classical gauge invariance leads to zero-norm states in the quantum
theory whenever there is an energy
level with energy $4M(N'+M)\,$, and we confirm this below. Note, in particular, that
this implies that there are zero norm ground states when $M=0\,$.

Before proceeding to the quantum theory we have to address a minor difficulty. The Hamiltonian $H_0$
does not commute, even `weakly',  with the constraints. This difficulty can be circumvented
by introducing the new variables
\be
\xi^1= \zeta + z\xi\, , \qquad \xi^2 = \xi\, . \lb{NewCo}
\ee
These were the variables used in \cite{Ivanov:2004yw}, and the analog of $H_0$ found by using these variables
commutes with the constraints. Alternatively, one can modify the Hamiltonian by adding terms proportional
to the constraint functions such that the new Hamiltonian commutes, at least weakly, with the constraints.
This second approach was the one adopted in \cite{Curtright:2006aq} for the planar superflag, and we will
do the same here. Specifically, we take the new Hamiltonian to be
\be
H = K_2^2K_1\left({\cal P}_z +i\xi{\cal P}_\zeta\right)\left({\cal P}_{\bar z} +i\xi{\cal P}_{\bar \zeta}\right).
\ee
It may be verified that $H$ is weakly equivalent to $H_0$ but commutes (strongly)  with the constraints.

\subsection{Quantum Theory}

To pass to the quantum theory we make the replacement $P_A\to -i \partial_A$, as in (\ref{quantP}),
where $A=(z,\zeta)$, and we also make the replacement
\be
\pi_\xi \to \partial_\xi\, , \qquad \pi_{\bar\xi} \to \partial_{\bar\xi}\, ,
\ee
which is needed only for  the second pair of constraints (\ref{secondpair}).
The resulting Hamiltonian operator\footnote{Operator ordering ambiguities allow the addition of a constant,
which we have set to zero.} is
\be
H_{N'} =   -K_2^2K_1 \left(\nabla_z^{(N')} -\xi\nabla_\zeta^{(N')}\right)\left(\nabla_{\bar z}^{(N')}
- {\bar \xi}\nabla_{\bar \zeta}^{(N')}\right),
\label{hat2}
\ee
where
\be\label{identity}
\nabla_A^{(N')} =  \partial_A - i N'{\cal A}_A\,, \qquad \nabla_{\bar A}^{(N')} =
\partial_{\bar A} - i N'{\cal A}_{\bar A}\, .
\ee
Because the analytic constraint operators commute, we may quantize {\it \`a la}
Gupta-Bleuler by requiring physical states to be annihilated by these operators. The result
is that  `physical' wave functions must take the form
\be\label{reduced}
\Psi = K_1^{M} K_2^{-N'} \Phi\left(z,\bar z_{sh},\zeta,\xi\right),
\ee
where $\Phi$ is a `reduced' wave function that depends on $\bar z$ only through the
 `shifted'  coordinate
\be\label{shifted}
\bar{z}_{sh} = \bar z - \xi\bar\zeta - \bar z\left(\zeta + z\xi\right)\bar\zeta\, .
\ee

For $2N'$ an integer, which we may assume to be positive, the Hamiltonian may be diagonalized
in the physical subspace, with energy eigenvalues
\cite{Ivanov:2004yw}
\be
E_{N'} =  \ell(2N' + \ell + 1)  \, , \qquad \ell=0,1,2,\dots\,.
\ee
The wave functions for the LLL ($\ell=0$) is
\be
\Psi^{(0)} =  K_1^{M} K_2^{-N'}\Phi^{(0)}_{an} \left( z,\zeta,\xi\right).
\ee
That is, the reduced LLL wave function is an {\it analytic} function. The reduced wave function
at all higher levels may be expressed in terms of a level $\ell$ {\it analytic} function
$\Phi_{an}^{(\ell)}$ according to the formula
\be\label{vectors}
\Phi^{(\ell)} =  {\cal D}^{2(N'+1)} \cdots {\cal D}^{2(N'+\ell)} \,
\Phi_{an}^{(\ell)}\left(z,\zeta,\xi\right) \qquad (\ell>0)\,,
\ee
where
\be
{\cal D}^{2N'} \equiv  \nabla_z^{2N'} -\xi\nabla_\zeta^{2N'} = {\partial_{z}}
- \xi\partial_\zeta -\frac{2N' \,\bar z_{sh}}{1 + z\bar z_{sh}}\, .
\ee

As in the case of the superspherical  Landau model, there is a natural $SU(2|1)$ invariant inner
product on Hilbert space defined by a superspace integral, although the superspace
now has an additional complex anti-commuting coordinate. As shown in \cite{Ivanov:2004yw}, this inner product  is
\be
\langle \Upsilon |\Psi\rangle = \int \! dz d\bar z \,
\partial_\zeta\partial_{\bar\zeta} \partial_\xi\partial_{\bar\xi} \, K_2^{-2}\, \Upsilon^* \Psi\, .\lb{meas2}
\ee
Performing the Berezin integration over {\it all} anti-commuting coordinates, we
get an ordinary integral over the sphere with an integrand determined by the four analytic functions
$(A^{(\ell)},\psi^{(\ell)},\chi^{(\ell)},F^{(\ell)})$ appearing in the
$(\zeta,\xi)$-expansion of  $\Phi_{an}^{(\ell)}\,$:
\be\label{anexp}
\Phi^{(\ell)}_{an} = A^{(\ell)}+ \zeta\left[ \psi^{(\ell)}+
\frac{\partial_z \chi^{(\ell)}}{\left(2N'+2\ell +1\right)} \right] + \xi\chi^{(\ell)} +
\zeta\xi \, F^{(\ell)}\, .
\ee
The net result, after integrating by parts to remove all derivatives, is that wave functions
at different levels are orthogonal, while
\bea\label{naivenorm}
||\Psi_{N'}^{(\ell)}||^2 &\equiv& \langle \Psi |\Psi\rangle = \ell! \frac{(2N'+\ell+1)!}{(2N'+1)!}
\int \frac{dz d\bar z}{\left(1+ z\bar z\right)^{2(N'+\ell+1)}} \times\nonumber\\
&&\Bigg\{ \left(2M-\ell\right) \left(2M+2N' +\ell +1\right) \bar A^{(\ell)} A^{(\ell)}
+ \bar F^{(\ell)} F^{(\ell)} \nonumber\\
&&+\
\frac{\left(N'+\ell +1\right)\left(2N'+2M +\ell +1\right)}{\left(2N'+2\ell +1\right)\left(1+ z\bar z\right)}
\bar\chi^{(\ell)} \chi^{(\ell)} \nn
&&+\  \left(2M-\ell\right)\left(1+ z\bar z\right) \bar\psi^{(\ell)} \psi ^{(\ell)}\Bigg\}\, .
\eea
This is a simplified form of the result given in \cite{Ivanov:2004yw}; the unusual expansion
of (\ref{anexp}) has led to a norm that is diagonal in the component functions. The finiteness
of the norm (the $S^2$ square-integrability requirement) requires, as usual, that fields of  $SU(2)$ spin $s$
are degree $2s$ holomorphic polynomials in $z$. The $SU(2)$ spin content will be computed explicitly
 in the following subsection, but it is not difficult to see what the result will be. The fields
 $A^{(\ell)}(z)$ and $F^{(\ell)}(z)$ each have spin $s = N' + \ell$, while
 the fields $\chi^{(\ell)}(z)$ and $\psi^{(\ell)}(z)$ have, respectively, spins
$s = N' + \ell +\frac{1}{2}$ and $s = N' + \ell -\frac{1}{2}$. This demonstrates, in particular,
the equality of the numbers of fermionic and bosonic degrees of freedom at any Landau
level without zero-norm states.

 With the above norm, the model has ghosts. For positive $M$ (which was the only case considered
in \cite{Ivanov:2004yw}) there are ghosts whenever $\ell>2M$ and if $2M$ is a non-negative integer
then there are zero-norm states for $\ell=2M\,$. This means, in particular, that the model
has ghosts in this `naive' norm for any positive $M\,$. The same is true for negative $M\,$,
and in this case  there are zero norm states even for $\ell=0\,$.

Of course, the sign of the norm has physical relevance only for Grassmann-even component functions,
and either $A^{(\ell)}$ or $\psi^{(\ell)}$ would be Grassmann-odd if we were to assume
(as in \cite{Ivanov:2004yw}) that  wave functions are superfields (i.e. have definite Grassmann parity).
However, even in this case the above statements concerning ghosts still apply. We have
been careful to allow  for (i) wave functions that are superfields, in which case the
`Hilbert' space is actually a vector superspace, and (ii)  wave functions
for which all component fields are ordinary functions  (or bundle sections),
in which case the Hilbert space is a vector space.  The ghost problem can be
circumvented by another choice of $SU(2|1)$-invariant norm, but we postpone
the construction of this alternative norm  until we have achieved
a better understanding of  the action of $SU(2|1)$ on the superflag Hilbert space.

\subsection{Unitary norm}

The $SU(2|1)$ symmetry of the superflag model implies the existence of Noether charges,
which become differential operators in the quantum theory, satisfying the (anti)commutation relations
of $SU(2|1)$ given in Section \ref{subsec:supersphere}. These differential operators acting on
the whole superflag wave functions, determine
a simpler set of differential operators that act on the analytic wave functions, and vice-versa since the full
Noether charge operators can be recovered from the simpler `analytic' operators that we
now present. The even generators are
\bea\label{evengen}
J_- &=&-i \partial_z\,,  \nonumber\\
 J_+  &=& -i\left[-2\left(N'+\ell\right) z + z^2\partial_z + z\zeta\partial_\zeta -
  \left(\zeta + z\xi\right)\partial_\xi \right], \nonumber\\
J_3 &=& -\left(N'+\ell\right) + z\partial_z + \frac{1}{2}\left(\zeta\partial_\zeta - \xi\partial_\xi\right),
\nonumber\\
F &=& 2M+N' + \frac{1}{2}\left(\zeta\partial_\zeta + \xi\partial_\xi\right)\, .
\eea
Note the $\ell$-independence of $B\,$; for the other generators one should view $\ell$
as an operator (later to be called $L$) that takes the value $\ell$ in the $\ell$th level.  The odd generators are
\be
\Pi = \partial_\zeta \, , \qquad Q= z\partial_\zeta - \partial_\xi
\ee
and
\bea
\Pi^\dagger &=& \left(2M+2N'+\ell\right)\zeta - \zeta z\partial_z +
\xi\left[\left(2M-\ell\right)z - \zeta\partial_\xi\right], \nonumber\\
Q^\dagger &=& \zeta \partial_z - \left(2M-\ell \right)\xi \, .
\eea
These results may be compared to the expressions  (\ref{ssgens}). In the present case, the full
differential operators representing the generators $(J_+,\Pi^\dagger, Q^\dagger)\,$,
which are determined by the simpler `analytic' forms given above,  are the Hermitian conjugates of the
generators  $(J_-,\Pi, Q)$ in the `naive'  norm.

We are  now in a position to work out  the $SU(2|1)$ representation content at each Landau level.
Let us first consider the $SU(2)$ content. We have
\bea
J^2 &=& J_-J_+ +J_3^2 +J_3 \nn
&=& \left(N'+\ell+1\right) \left(N'+\ell\right) - \left(N'+\ell + \frac{1}{4}\right) \zeta\partial_\zeta \nn
&&+\ \left[\zeta\partial_z + \left(N'+\ell+ \frac{3}{4} -
\frac{1}{2}\zeta\partial_\zeta\right)\xi\right]\partial_\xi\,.
\eea
Now we act with this operator on the analytic wave functions of (\ref{anexp}), which we may rewrite as
\be
\Phi^{(\ell)}_{an} = A^{(\ell)}+ \zeta \psi^{(\ell)}+
\left[ \xi+ \frac{\zeta\partial_z}{2N'+2\ell+1}\right]\chi^{(\ell)} + \zeta\xi \, F^{(\ell)}\, .
\ee
We find that
\bea
J^2 \Phi^{(\ell)}_{an} &=& \left(N'+\ell\right)\left(N'+\ell +1\right) A^{(\ell)}
+ \left(N'+\ell -\frac{1}{2}\right)\left(N'+\ell+\frac{1}{2}\right) \zeta\psi^{(\ell)} \nn
&&+\, \left(N'+\ell + \frac{1}{2}\right)\left(N'+\ell+\frac{3}{2}\right)
\left[ \xi +  \frac{\zeta\partial_z}{2N'+2\ell+1}\right]\chi ^{(\ell)} \nn
&&+ \left(N'+\ell\right)\left(N'+\ell +1\right)  \zeta\xi\, F^{(\ell)}\, .
\eea
One reads off from this result the eigenfunctions of $J^2$ and their eigenvalues.
Acting with $J_3$ on the $J^2$ eigenfunctions we get
\bea
J_3 \left[A^{(\ell)}\right] &=& \left(z\partial_z -N'-\ell\right)A^{(\ell)}\, , \nn
J_3\left[ \zeta\psi^{(\ell)}\right] &=& \zeta \left(z\partial_z -N'-\ell + \frac{1}{2}\right)\psi^{(\ell)}\,,\nn
J_3\left[ \left(\xi + \frac{\zeta\partial_z}{2N'+2\ell+1}\right)\chi^{(\ell)}\right] &=&
 \left(\xi + \frac{\zeta\partial_z}{2N'+2\ell+1}\right)\left(z\partial_z -N'-\ell
 -\frac{1}{2}\right)\chi^{(\ell)} \nn
J_3 \left[\zeta\xi\, F^{(\ell)}\right] &=& \zeta\xi \left(z\partial_z -N'-\ell\right) F^{(\ell)}\,.
\eea
Putting this all together we find the following  sets of  $(2s+1)$ spin-$s$
joint eigenfunctions of $J^2$ and $J_3\,$:
\bea
s&=& \left(N'+\ell\right)\, :\  z^na_n\, , \qquad n=0, \dots, 2N'+2\ell\,, \nn
s&=& \left(N'+\ell -\frac{1}{2}\right)\, : \ \zeta z^p \psi_n \, , \qquad p=0,\dots, 2N'+2\ell-1 \,,\nn
s&=& \left(N'+\ell +\frac{1}{2}\right)\ :
\left(\xi + \frac{\left(q+1\right)\zeta}{2N'+2\ell+1}\right) z^q\chi_q
\, , \qquad q=0, \dots, 2N'+2\ell+1\,,\nn
s&=& \left(N'+\ell\right)\, : \ \zeta\xi\, z^mf_m\, , \qquad m=0, \dots, 2N'+2\ell
\eea
for constants  $(a_m, \psi_p,\chi_q,f_m)\,$.

As mentioned already, there are two separate cases in which the `naive' norm considered
so far has ghosts when $M<0$. These are (i) $2M< -2N'-1\,$, and (ii) $-2N'-1< 2M <0\,$.
Consider the operator
\be\label{metricop1}
G_{an}= -1 + 2\xi\partial_\xi + \frac{2}{2N+2\ell+1}\, \zeta\partial_z\partial_\xi\, .
\ee
This commutes with $J^2$ and $J_3$, and hence with the Hamiltonian,  as is clear
from the alternative expression
\be
G_{an}= \frac{1}{2N' + 2\ell+1} \left[2J^2 + 2\left(F -2M +\ell\right)^2
- \left(2N'+ 2 \ell +1\right)^2\right].
\ee
It also  has the property that
\be
G_{an}^2\equiv 1\, .
\ee
As explained in \cite{Curtright:2006aq}, the same properties hold for the corresponding `full'
operator $G$,  so each of the eigenstates listed above has a definite  `$G$-parity'.
By inspection, one sees that for
\be\label{range1}
-2N' -1  < 2M < 0 \, ,
\ee
the positive (negative) norm eigenstates have positive (negative) $G\,$-parity, and therefore that
the $G$ is  the `metric operator' for $M$ in the above range, in the sense that the new norm
\be
||| \Psi|||^2 \equiv  \langle \Psi | G\Psi\rangle
\ee
is positive definite; we refer to  \cite{Curtright:2006aq} for details of the formalism.
In the planar limit, this range extends to all negative $M\,$, so we should expect the planar limit
of $G_{an}$ to be the $M<0$ metric operator of the planar superflag found in \cite{Curtright:2006aq},
and this is indeed the case. For $M=0$ there are zero-norm states, as in the planar limit, but still no negative-norm states.
This allows us to redefine the states in a `physical' Hilbert space to be
equivalence classes of states  in the original Hilbert space in which two states that differ by a
zero-norm state are considered equivalent.

Now consider the operator
\be
\tilde G_{an}= 1- 8\left(F-2M-N'\right) +8\left(F-2M-N'\right)^2 \, .
\ee
It is manifest that $\tilde G_{an}$ commutes with the Hamiltonian, and hence the same is true
of $\tilde G$.
One may verify that $\tilde G_{an}^2\equiv 1\,$, so that the eigenstates listed above also have a
definite  $\tilde G\,$-parity. Inspection shows that when $2M< -2N'-1$ the states with positive
(negative) norm have (positive) negative $\tilde G$-parity. The operator $\tilde G$
is therefore a `metric' operator for $2M< -2N'-1\,$, which is a range that has
no counterpart in the planar limit. As in the planar limit, the metric operator for $M>0$
is a more-complicated `dynamical' one, depending on the level. We skip the details of this case.

%%%%%%%%%%%%%%%%%%%%%%%%%%%%%%
\section{Hidden symmetries}
\setcounter{equation}{0}

We know that there is hidden worldline supersymmetry of the {\it planar} super-Landau models,
for $M\le 0\,$. This implies the existence of some enlarged supersymmetry algebra for the spherical
super-Landau models, and we now aim to investigate this. For simplicity, we now place $M$ in the range
for which the metric operator defining the unitary models is the operator $G$ defined by  (\ref{metricop1}).
As we have seen, this means that $M$ should satisfy (\ref{range1}) but, as we have also seen,
we may allow $M=0$ too.  In other words, we now restrict $M$ such that
\be
 -2N' -1< 2M \le 0 \, .
 \ee
Now, let ${\cal O}$ be some operator that commutes with the Hamiltonian, and hence generates some symmetry
of the model under investigation, and let ${\cal O}^\dagger$ be its hermitian conjugate with respect
to the `naive', and non-positive, Hilbert space norm. Then its hermitian conjugate with respect
to the positive Hilbert space norm  is (recall that $G^2\equiv 1$)
\be
{\cal O}^\ddagger \equiv G {\cal O}^\dagger G  =
{\cal O}^\dagger + G{\cal O}_G^\dagger\, ,
\ee
where
\be
{\cal O}_G \equiv \left[G,{\cal O}\right]
\ee
is another operator that commutes with the Hamiltonian. Note that
\be
  \left({\cal O}_G\right)^\ddagger = \left[G,{\cal O}^\dagger\right] =
-\left[G,{\cal O}\right]^\dagger = - \left({\cal O}_G\right)^\dagger \equiv - {\cal O}_G^\dagger\, .
\ee
Symmetry generators that do not commute with $G$ thus generate, in general, additional
symmetries that are `hidden' in the sense that their existence was not built into the construction
of the model.  For the superflag model, it is the odd generators that fail to commute with $G$,
and this leads to the following new symmetry generators
\bea
\Pi_G&=& -\frac{2}{2N'+2\ell + 1}\, \partial_\xi \partial_z\,, \nn
\Pi_G^\ddagger  &=&\frac{4M-2\ell}{2N'+2\ell+ 1}\left[\zeta\left(1+ z\partial_z\right)
+ \left(2N'+2\ell +1\right)z\xi - \zeta \xi\partial_\xi  \right],\nn
Q_G&=&\frac{2}{2N'+2\ell + 1} \left(2N' + 2\ell + 1 -z\partial_z -\zeta\partial_\zeta\right)\partial_\xi\,,  \nn
Q_G^\ddagger&=& -\frac{4M-2\ell}{2N'+2\ell+ 1}\left[\left(2N'+2\ell+ 1\right)\xi + \zeta\partial_z \right].  \label{dgen}
\eea

The naive hermitian conjugate of a symmetry operator ${\cal O}$ will not coincide with its new hermitian conjugate
${\cal O}^\ddagger$ unless  ${\cal O}$ commutes with $G\,$. For this reason, it  is convenient
to choose a basis  in which the original $SU(2|1)$ symmetry operators ${\cal O}$ are replaced by the operators
\be\label{tildedef}
\tilde{\cal O}  = {\cal O} + \frac{1}{2}{\cal O}_G G \, ,
\ee
which commute with $G$ even when ${\cal O}$ does not. This property means that
\be
\tilde{\cal O}^\ddagger = \tilde{\cal O}^\dagger = {\cal O}^\dagger - \frac{1}{2}G{\cal O}^\ddagger_G\, .
\ee
In the case that  ${\cal O}$ is hermitian with respect to the `naive' Hilbert space metric, the operator
$\tilde{\cal O}$ will  be hermitian with respect to the new Hilbert space norm.

When applied to the operators $\Pi$ and $Q$, the definition (\ref{tildedef}) yields
\bea
\tilde \Pi &=& \Pi + \frac{1}{2}\Pi_G \, , \qquad
\tilde \Pi^\dagger = \Pi^\dagger - \frac{1}{2}\Pi^\ddagger_G \, , \nn
\tilde Q &=& Q+ \frac{1}{2}Q_G \, , \qquad
\tilde Q^\dagger = Q^\dagger - \frac{1}{2}Q^\ddagger_G \, ,
\eea
where we have used the remarkable identities
\be
\Pi_G G = \Pi_G\, , \qquad  Q_GG = Q_G\, .
\ee
In terms of the rescaled odd charges
\be
\left(\tilde \Pi' , \tilde Q'  \right) = \sqrt{ \frac{2N'+2\ell+ 1}{2M+2N' +\ell+ 1}}
\left(\tilde \Pi,  \tilde Q \right)\, ,
\ee
and the redefined $U(1)$ generator
\be
F' = F-2M+ \ell \, ,
\ee
one finds, after some computation,  that the non-zero (anti)commutation relations of the odd charges
$(\tilde \Pi', \tilde Q')\,$, and their hermitian conjugates,  and the even $SU(2)\times U(1)$ charges
$(J_3,J_\pm, F')$ are  precisely of the standard  $SU(2|1)$ form given in Section \ref{subsec:supersphere}.
Thus, these charges provide an alternative basis for the $SU(2|1)$
symmetry algebra.

Now we turn to the  `hidden' symmetry charges.   Their non-zero anticommutators are
\bea\label{sl21_G}
\left \{\Pi_G, \Pi_G^\ddagger \right\} &=&  \frac{4\left(\ell-2M\right)}{2N'+ 2\ell +1}
 \left(J_3 + \check F\right),
\qquad  \left \{Q_G, Q_G^\ddagger\right\} =  \frac{4\left(\ell-2M\right)}{2N'+2\ell + 1}\left(- J_3 + \check F\right), \nn
\left\{\Pi_G,  Q_G^\ddagger\right\} &=& -i\frac{4\left(\ell-2M\right)}{2N'+2\ell + 1} \, J_-\, , \qquad
\left\{\Pi_G^\ddagger, Q_G\right\} = i\frac{4\left(\ell-2M\right)}{2N'+2\ell + 1} \, J_+\, ,
\eea
where
\be\label{checkB}
\check F=  2M + 2N' +\ell + 1 -F\, .
\ee
Notice that  the coefficients are level-dependent.  The $\ell$-dependence in the denominators
is easily removed by a level-dependent rescaling of the odd charges but the $(\ell-2M)$ factor
in the numerators is more problematic because when $M=0$ this factor is zero for $\ell=0$ but non-zero
for $\ell>0\,$. For this reason, we will discuss these two cases separately.

\subsection{$-2N'-1<2M<0$}

In this case we may define new odd charges by
\be
\check\Pi_G = - \sqrt{\frac{2N'+2\ell + 1}{4\left(\ell -2M\right)}}\, Q_G^\ddagger\, , \qquad
\check Q_G = \sqrt{\frac{2N'+2\ell + 1}{4\left(\ell -2M\right)}} \, \Pi_G^\ddagger \, ,
\ee
in terms of which the anti-commutation relations  of (\ref{sl21_G})  become
\bea
\left\{\check\Pi_G,\check\Pi_G^\ddagger\right\} &=& -J_3 + \check F \, , \qquad
\left\{\check Q_G,\check Q_G^\ddagger\right\} = J_3 + \check F\,,  \nonumber\\
\left\{\check\Pi_G, \check Q_G^\ddagger\right\} &=& iJ_- \, , \qquad
\left\{\check \Pi_G^\ddagger,\check Q_G\right\} = -iJ_+\, .
\eea
To present  the commutators of these new odd charges with the even charges of $SU(2|1)$
we need give only the non-zero commutators with $(\check\Pi_G,\check Q_G)$ charges since the remainder
are found by hermitian conjugation; these are
\bea
\left[\check F, \check\Pi_G\right] &=& -\frac{1}{2}\check \Pi_G \,, \qquad
 \left[\check F, \check Q_G\right] = -\frac{1}{2}\check Q_G\,,\nn
\left[J_3,  \check \Pi_G\right]  &=& -\frac{1}{2}\check \Pi_G\, , \qquad
 \left[J_3, \check Q_G\right] =  \frac{1}{2}\check Q_G\,, \nn
 \left[ J_+, \check \Pi_G\right] & = & i \check Q_G\,, \qquad
 \left[J_-, \check Q_G\right] = -i\check\Pi_G\, .
\eea
This shows that the new odd symmetry charges transform as a charged doublet under the $U(2)$ subgroup of  $SU(2|1)$.
In fact, the operators  $(\check\Pi^\ddagger, \check Q^\ddagger)$, together with their hermitian conjugates,
and the  even charges $(J_3,J_\pm, \check F)$, obey the (anti)commutation relations  of $SU(2|1)$ given
in (\ref{subsec:supersphere}). The full symmetry group therefore contains two  distinct  $SU(2|1)$ superalgebras.
 As $F'$ is the $U(1)$ charge of one
of these superalgebras and $\check F$ the $U(1)$ charge of the other one, the full symmetry group
must contain
\be
Z=F' + \check F = 2N' +2\ell+1\, ,
\ee
which is a level-dependent central charge. However,  this level-dependence does not present a
problem; it just means that we have a central charge
\be
Z= 2L+ 2N'+1\,,\lb{CCharg}
\ee
where $L$ is the level operator .

The two $SU(2|1)$ superalgebras are non-commuting because there are non-zero  anti-commutators
of  the odd charges from one with the odd charges from the other. These are
\bea
\left\{\tilde\Pi',\check \Pi_G^\ddagger \right\} &=&
\left\{\tilde Q',\check Q_G^\ddagger \right\} = i{\cal J}_-\,,
\nonumber\\
\left\{\tilde\Pi'{}^\ddagger,\check \Pi _G\right\} &=&
\left\{\tilde Q'{}^\ddagger,\check Q_G\right\} = -i{\cal J}_+\, ,
\eea
where the analytic operators representing ${\cal J}_\pm$ are
\bea
{\cal J}_+ &=& i \sqrt{\left(\ell-2M\right)\left(2M+2N'+\ell+1\right)}\, \xi\zeta\, , \nn
{\cal J}_-  &=& \frac{i}{\sqrt{\left(\ell-2M\right)\left(2M+2N'+\ell+1\right)}}\, \partial_\xi\partial_\zeta \, .
\eea
These satisfy, together with
\be
{\cal J}_3 = \frac{1}{2}\left(-1+ \xi\partial_\xi + \zeta\partial_\zeta\right),
\ee
the  standard $su(2)$ commutation relations
\be
\left[{\cal J}_+,{\cal J}_-\right] = 2{\cal J}_3\, , \qquad
\left[{\cal J}_3,{\cal J}_\pm\right] = \pm {\cal J}_\pm\, .
\ee

Finally, the non-zero commutators of these new $SU(2)$ charges with the odd charges are
\bea
\left[ {\cal J}_+, {\tilde \Pi}^\prime\right]  &=&- i\check \Pi_G, \qquad \quad
\left[ {\cal J}_+, {\tilde Q}^\prime\right] =  i \check Q_G\, , \nn
\left[ {\cal J}_-, {\Pi}^\prime_G\right]  &=&  i{\tilde \Pi}^\prime, \qquad \quad
\left[ {\cal J}_-, \check Q_G\right] =- i {\tilde Q}^\prime\, , \nn
 \left[{\cal J}_3, {\tilde \Pi}^\prime\right] &=& -\frac{1}{2}{\tilde \Pi}^\prime\, , \qquad
 \left[{\cal J}_3, {\tilde Q}^\prime\right] = -\frac{1}{2}{\tilde Q}^\prime  \, , \nn
 \left[ {\cal J}_3, \check \Pi_G\right] &=& \frac{1}{2}\check \Pi_G\, , \qquad \quad
\left[  {\cal J}_3, \check Q_G\right]  = \frac{1}{2}\check Q_G\,  ,
\eea
and hermitian conjugates. These commutation relations show that $(\tilde \Pi', \check \Pi_G)$ and
$(\tilde Q',\check Q_G)$ are doublets of the $SU(2)$ group generated by $({\cal J}_\pm,{\cal J}_3)\,$.

We have now shown that the charges
\be
\{J_\pm,J_3,{\cal J}_\pm,{\cal J}_3, Z;
\tilde\Pi^\prime, \tilde Q^\prime; \check \Pi_G,\check Q_G\}
\ee
span a  Lie superalgebra, with structure constants that are level independent. We have therefore found
 a finite-dimensional `enlarged' symmetry algebra. The brackets where the central charge $Z$
 defined in \p{CCharg} contributes, are:
 \bea \label{oddgen}
 \{ {\check\Pi_G}, {\check \Pi}_G^\ddagger \} =  -J_3 - {\cal J}_3 + Z\,, \;
 \{{\check Q}_G, {{\check Q}_G}^\ddagger\}
 =  J_3 - {\cal J}_3 + Z  \,,
\; \nn
\{{\tilde \Pi^\prime}, {\tilde {\Pi^\prime}}^\dagger \} =  -J_3 + {\cal J}_3 + Z\,,
\; \{{\tilde Q^\prime}, {\tilde{ Q^\prime}}^\dagger\} =  J_3 + {\cal J}_3 +  Z\,.
\;
\eea
Its even subalgebra is that of
 $SU(2)\times SU(2) \times U(1)$, where the $U(1)$ charge is central, and its  four complex
 odd generators transform as the
 $\left({\bf 2},{\bf 1}\right) \oplus \left({\bf 1},{\bf 2}\right)$ of  $SU(2)\times SU(2)$.
 This uniquely fixes the full symmetry algebra to be that of $SU(2|2)$; recall that the groups $SU(p|q)$
 have even subgroup $SU(p)\times SU(q)\times U(1)$ with the $U(1)$ charge being central when
 $p=q\,$.

\subsubsection{Casimir considerations}

Acting on the wave functions at the $\ell$th level, the $SU(2|1)$ Casimir operators \p{Cas2} and \p{Cas3}
for the superflag model become
\be
C_2 = \left(\ell-2M\right)\left(2M+2N'+\ell +1\right), \qquad
C_3= \left(4M + 2N' +1\right)C_2\, .
\ee
For the general superflag model, one has
\be
H= C_2 + 2M\left(2M+2N'+1\right). \lb{HSF11}
\ee
At levels for which $C_2=0\,$, which is possible when $2M$ is a non-negative integer, then $C_3=0$ too,
and hence the $SU(2|1)$ representation is `atypical'.  In particular, $C_2=C_3=0$ for the LLL when
$M=0$, in which case
\be
H=C_2\big|_{M=0} = \ell\left(\ell + 2N'+1\right) \, ,
\ee
in agreement with our result of (\ref{casfunction}) for the  supersphere if we make the identification
\be\label{NprimeN}
2N'= 2N-1\, .
\ee

The $su(2|2)$ symmetry algebra for $M<0$ is a subalgebra of the enveloping algebra of $su(2|1)\,$.
To see this we define the  following functions of the Casimir operators:
\bea
{\cal A} &=& \sqrt{{C_3 + \sqrt{C_3^2 +4C_2^3} \over 2C_2\ \sqrt{C_3^2 +4C_2^3}}}\, , \nn
{\cal B}  &=& \sqrt{{1 \over 2 C_2^3}\left( C_3^2 +4C_2^3 - C_3 \sqrt{C_3^2 +4C_2^3}\right)}\, , \nn
{\cal C}  &=& \sqrt{{1 \over 2 C_2^3}\left( C_3^2 +2C_2^3 - C_3 \sqrt{C_3^2 +4C_2^3}\right)}\, .
\eea
The odd charges of $SU(2|2)$  may now be written as
\bea
\check \Pi_G &=& {\cal A}\left[iJ_-\Pi^\dagger  +Q^\dagger\left(J_3 -F+{C_3 \over 2 C_2}
- \sqrt{\left({C_3 \over 2 C_2} \right)^2+ C_2}  \right)\right],\nn
\check Q_G &=& {\cal A}\left[iQ^\dagger J_+ +\Pi^\dagger\left(J_3 +F+1 {-{C_3 \over 2 C_2}}
+ \sqrt{{\left(C_3 \over 2 C_2\right)^2} + C_2} \right)\right], \nn
 {\tilde \Pi}^\prime &=& {\cal B} \Pi +  {\cal C} \check Q_G^ \ddagger \, ,\qquad
 {\tilde Q}^\prime= {\cal B}Q - {\cal C}\check \Pi_G^\ddagger \, .
 \eea
 The even charges are those of the original $SU(2)$ symmetry, $(J_\pm,J_3)\,$, the central charge
 $Z=L+2N+1\,$, and the `hidden' $SU(2)$ charges
\bea
{\cal J}_- ={i \over \sqrt{C_2}}\,\Pi Q\,,\qquad  {\cal J}_+
={i \sqrt{C_2}}\, Q^\dagger \Pi^\dagger,\qquad {\mathcal J}_3 = F - \frac{C_3}{2C_2}\, .
\eea

\subsection{$M=0$ and the planar limit}

In this case the anticommutation relations (\ref{sl21_G})  reduce to
\bea\label{sl21_Greduced}
 \{\Pi_G, \Pi_G^\ddagger \} &=&  \frac{4\ell}{2N+ 2\ell } \left(J_3 + \check F\right), \nn
  \{Q_G, Q_G^\ddagger\} &=&  \frac{4\ell}{2N+2\ell }\left(- J_3 + \check F\right), \nn
\{\Pi_G,  Q_G^\ddagger\} &=& -i\frac{4\ell}{2N+2\ell } J_-\, ,
\eea
where we have used (\ref{NprimeN}).   As all these anti-commutators vanish for $\ell=0$,
the LLL states must be annihilated by {\it both} $(\Pi_G,Q_G)$ and their hermitian conjugates
$(\Pi_G^\ddagger,Q_G^\ddagger)\,$.  At higher levels, we get supermultiplets of states
that may be constructed by the repeated action of $(\Pi_G^\ddagger,Q_G^\ddagger)$ on
`Clifford  vacuum'  states annihilated  by $(\Pi_G,Q_G)\,$. In fact,  all higher levels
may be shown to form representations of $SU(2|2)$ by the argument
just used to analyze all levels when $M<0\,$.  However, because of the exceptional  LLL
for $M=0$, one cannot say that the model has an $SU(2|2)$ symmetry. Neither is there
a conventional supersymmetry, as there is in the planar limit, because the commutators of the
`supersymmetry' generators $(\Pi_G,Q_G)$ with the even generators of $SU(2|1)$ produce
further odd symmetry generators. In fact,  closure of the algebra appears to require an infinite
number  of generators.

As this state of affairs is in marked contrast to the simple results obtained in \cite{Curtright:2006aq}
for the superplane Landau model, we now discuss how those results may be recovered in the planar limit.
To do so we must restore dependence on the radius $R$ of the sphere that is the `body'
of both the supersphere and superflag supermanifolds. Specifically, the Hamiltonian must be rescaled:
\be
H\to H/R^2 = 2\ell \left(N/R^2 + \ell/R^2\right)\, .
\ee
We then take $R\to\infty\,$, keeping fixed
\be
\kappa = N/R^2\, .
\ee
This gives
\be
H_{superplane} = 2\kappa L\,,
\ee
where $L$ is the level operator with eigenvalue $\ell$ on the $\ell$ th level. This agrees with
 \cite{Ivanov:2005vh,Curtright:2006aq} after taking into account the difference in notations
 of that paper\footnote{Confusingly for present purposes, the  level number  $\ell$ was
 called $N$ in  \cite{Ivanov:2005vh,Curtright:2006aq}. The parameter $N$ used here does not appear as such in
 the planar limit because it is replaced by the real number $\kappa\,$.}.

{}From the $N$ dependence of the generators $(J_\pm,J_3,\check F)$ we find that
\bea
J_-/R^2 = {\cal O}\left(1/R^2\right), \qquad && J_+ =  -2i \kappa z + {\cal O}\left(1/R^2\right),
\nn
\check F + J_3 = {\cal O}\left(1/R^2\right), \qquad &&
\check F -J_3 = 2\kappa + {\cal O}\left(1/R^2\right).
\eea
The anti-commutation relations (\ref{sl21_Greduced}) can now be written as
\be
 \{\Pi_G, \Pi_G^\ddagger \} =  {\cal O}\left(1/R^2\right), \qquad
 \{\Pi_G,  Q_G^\ddagger\} = {\cal O}\left(1/R^2\right),
 \ee
 and
 \be
  \{Q_G, Q_G^\ddagger\}  = 2\ell +  {\cal O}\left(1/R^2\right).
 \ee
Thus, only $Q_G$ survives the planar limit, and it is proportional to the worldline supersymmetry charge
$S$ of \cite{Curtright:2006aq}.

%%%%%%%%%%%%%%%%%%%%%%%%%%%%
\setcounter{equation}{0}
\section{Supersphere from superflag}

In this Section we show how the quantum states of the supersphere model and its Hamiltonian
can be recovered using the basic geometric objects of the superflag manifold $SU(2|1)/[U(1)\times U(1)]$.
The supersphere $SU(2|1)/U(1|1)$ is an $SU(2|1)$ invariant subspace in $SU(2|1)/[U(1)\times U(1)]$,
whence it follows that any considerations related to the supersphere should have
an equivalent formulation in terms of the properly constrained objects defined on the superflag.
Throughout this Section we assume that all wave functions are superfields, i.e. that they
have definite Grassmann parity.

\subsection{Covariant derivatives}

As shown in \cite{Ivanov:2004yw}, the geometry of the superflag manifold $SU(2|1)/[U(1)\times U(1)]$
is described by a set of covariant derivatives with non-trivial $U(1)\times U(1)$ connections:
\bea
&&{\cal D}^{-} = D^{-} - (D^- \log K_2)\,\hat{J}_3 + (D^- \log K_1)\, \hat{B}\, \,, \quad
{\cal D}^+ = D^+ + (D^+ \log K_1) \hat{B}\nn
&&\bar{\cal D}{}^- = \bar{D}^- - (\bar{D}^- \log K_1) \hat{B}\,, \quad
\bar{\cal D}{}^+ = \bar{D}^+ + (\bar{D}^+ \log K_2)\, \hat{J}_3
-(\bar{D}^+ \log K_1)\, \hat{B}\,, \nn
&& {\cal D}^{--} = D^{--} - (D^{--} \log K_2)\,\hat{J}_3\,, \quad
{\cal D}^{++} = D^{++} + (D^{++} \log K_2)\,\hat{J}_3 \, , \lb{CovSF1}
\eea
subject to the conjugation rules\footnote{These rules are the same as those for  the purely
`derivative' parts of the covariant derivatives.}
\be
\bar{\cal D}^{+} = \overline{({\cal D}^{-})}\,, \qquad \bar{\cal D}^{-} = \overline{({\cal D}^{+})}\,,
\qquad {\cal D}^{++} = \overline{({\cal D}^{--})}\, .
\ee
Explicit expressions for the covariant derivatives were given in \cite{Ivanov:2004yw}
for  local superflag coordinates $(z,\xi^1,\xi^2)$, where $(\xi^1,\xi^2)= (\zeta + z\xi, \xi)$. Here we
use the local coordinates $(z,\zeta,\xi)$, in which case
\bea
&& D^{-}= (K_1K_2)^{\frac{1}{2}}\left\{\partial_\zeta -K_2^{-1}\left(\bar z - \xi \bar\zeta \right)\partial_{\xi}
+ K_1^{-1}\left[(1 + z\bar z)\bar\xi + z\bar\zeta\right]\left(\partial_z - \xi\partial_\zeta \right)\right\}, \nn
&& \bar{D}^+ =
-(K_1K_2)^{\frac{1}{2}}\left\{\partial_{\bar\zeta} -K_2^{-1}\left(z + \bar\xi \zeta \right)\partial_{\bar\xi}
- K_1^{-1}\left[(1 + z\bar z)\xi + \bar z \zeta\right]\left(\partial_{\bar z} - \bar\xi\partial_{\bar\zeta} \right)\right\},\nn
&& D^+ = K_2^{-\frac{1}{2}}\partial_{\xi}\,,  \quad \bar{D}^- = -K_2^{-\frac{1}{2}}\partial_{\bar\xi}\,,  \nn
&& D^{--} = K_1^{\frac{1}{2}} K_2\left(\partial_z -\xi\partial_\zeta \right), \quad
D^{++} = K_1^{\frac{1}{2}} K_2\left(\partial_{\bar z} -\bar\xi\partial_{\bar\zeta} \right). \lb{CovSF2}
\eea
The functions $K_1$ and $K_2$ are given for our choice of coordinates\footnote{The expressions in \cite{Ivanov:2004yw}
differ because of the different coordinates used there.}  in  (\ref{Ks}). In \p{CovSF1}, the operators $\hat{B}$, $\hat{J}_3$ are
`matrix' parts of the $U(1)$ generators $J_3$ and $B\,$, where $B$ is related to the generator $F$ of the previous Sections by
\be
B = \frac{1}{2}\left(F - J_3\right).
\ee

The covariant derivatives have the following commutation relations with the operator $\hat{F}\,$:
\be
\left[\hat{F}\,, {\cal D}^{\pm}\right] = \frac{1}{2}\,{\cal D}^{\pm}\,, \qquad
\left[\hat{F}\,, \bar{\cal D}^{\pm}\right] =
-\frac{1}{2}\,\bar{\cal D}^{\pm}\,, \qquad
\left[\hat{F}\,, {\cal D}^{\pm\pm}\right] = 0\,. \lb{FBD}
\ee
It is also useful to have the commutation relations with the operators $\hat{J}_3$ and $\hat{B}\,$:
\bea
&& [\hat{J}_3\,, {\cal D}^{\pm}] = \pm \frac{1}{2}\,{\cal D}^{\pm}\,, \quad [\hat{J}_3\,, \bar{\cal D}^{\pm}]
= \pm \frac{1}{2}\,\bar{\cal D}^{\pm}\,, \quad
[\hat{J}_3\,, {\cal D}^{\pm\pm}] = \pm {\cal D}^{\pm\pm}\,, \nn
&& [\hat{B}\,, {\cal D}^{\pm\pm}] = \mp \frac{1}{2}\,{\cal D}^{\pm\pm}\,, \; [\hat{B}\,, {\cal D}^{-}] = \frac{1}{2}\,{\cal D}^{-}\,, \;
[\hat{B}\,,\bar{\cal D}^{+}] = -\frac{1}{2}\,\bar{\cal D}^{+}\,,   \nn
&& [\hat{B}\,,{\cal D}^{+}] = [\hat{B}\,,\bar{\cal D}^{-}] = 0\,. \lb{JBD}
\eea

In what follows, a crucial role will be played by the (anti)commutation relations
between the covariant derivatives:
\bea
&& \{{\cal D}^{-}, \bar{\cal D}^{-} \} = - {\cal D}^{--}\,, \quad
\{{\cal D}^{+}, \bar{\cal D}^{+} \} =  {\cal D}^{++}\,, \lb{SpinCom0}\nn
&& \{{\cal D}^{-}, \bar{\cal D}^{+} \} = 2(\hat{B} + \hat{J}_3) = \hat{F} + \hat{J}_3\,, \quad
\{{\cal D}^{+}, \bar{\cal D}^{-} \} = 2\hat{B} = \hat{F} - \hat{J}_3\,, \lb{SpinCom1} \nn
&& \{{\cal D}^{-}, {\cal D}^{+} \} =  \{\bar{\cal D}^{-}, \bar{\cal D}^{+} \}
= \{{\cal D}^{\pm}, {\cal D}^{\pm} \} = \{\bar{\cal D}^{\pm}, \bar{\cal D}^{\pm} \} = 0\,, \lb{SpinCom} \nn
&& [{\cal D}^{++}, {\cal D}^{-}] =  -{\cal D}^{+}, \, [{\cal D}^{++}, {\cal D}^{+}] = 0\,, \,
[{\cal D}^{++}, \bar{\cal D}^{-}] =  \bar{\cal D}^{+}, \, [{\cal D}^{++}, \bar{\cal D}^{+}] = 0\,, \nn
&& [{\cal D}^{--}, {\cal D}^{+}] =  {\cal D}^{-}, \, [{\cal D}^{--}, {\cal D}^{-}] = 0\,, \,
[{\cal D}^{--}, \bar{\cal D}^{+}] =  -\bar{\cal D}^{-}, \, [{\cal D}^{--}, \bar{\cal D}^{-}] = 0\,, \nn
&& [{\cal D}^{++}, {\cal D}^{--}] = -2\,\hat{J}_3\,. \lb{Comm2}
\eea
These relations are equivalent to the Maurer-Cartan equations for the left-invariant 1-forms
on the superflag $SU(2|1)/[U(1)\times U(1)]$, and so fully encode the geometry of this supercoset manifold.
They can be derived from the Maurer-Cartan equations on  the superflag manifold
without reference to the explicit form of the covariant derivatives.

\subsubsection{Superflag superfields}

The $U(1)\times U(1)$ operators $(\hat J_3, \hat B)$ have eigenvalues $(\hat N, \hat M)$. Let $\Psi^{(\hat N, \hat M)}$
denote an eigenfunction of these operators:
\be
\hat{J}_3\,\Psi^{(\hat N, \hat M)} = \hat N\, \Psi^{(\hat N, \hat M)}\,, \qquad
\hat{B}\, \Psi^{(\hat N, \hat M)} = \hat M\, \Psi^{(\hat N, \hat M)}\, . \lb{JBpsi}
\ee
A covariant derivative of any such eigenfunction (which is a superfield on the superflag manifold)
is another eigenfunction because the covariant derivatives have definite $U(1)\times U(1)$ charges
as a consequence of the commutation relations \p{FBD} and \p{JBD}.

The general $SU(2|1)/[U(1)\times U(1)]$ superfields $\Psi^{(\hat N, \hat M)}$ have the following transformation law under
the odd $SU(2|1)$ transformations  \cite{Ivanov:2004yw}:
\bea
\delta\Psi^{(\hat N, \hat M)} &=& - \hat N \left(\epsilon^1\bar\zeta + \bar\epsilon_1\zeta \right) \Psi^{(\hat N, \hat M)} \nn
&&-\, \hat M\left[\epsilon^1(\bar\zeta + \bar z\bar\xi) +
\epsilon^2\bar\xi + \bar\epsilon_1(\zeta + z\xi) +  \bar\epsilon_2 \xi\right] \Psi^{(\hat N, \hat M)}\,.
\lb{GenTranSF}
\eea

It should be appreciated that the $SU(2|1)/[U(1)\times U(1)]$ superfields defined by \p{JBpsi} are purely geometric objects
having no {\it a priori} relation to the quantum superflag or supersphere wave superfunctions that we
discussed in the previous Sections. Consequently,  the real eigenvalues $(\hat N, \hat M)$ are not obliged
to coincide with the model parameters $N, N'$ and $M$ appearing in the Lagrangians \p{Lag1}
and \p{LagSF}. Nevertheless, it will turn out that  the wave functions of the quantum superflag Landau model
are superfields $\tilde{\Psi}^{(\hat N, \hat M)}$ with $\hat N = N', \hat M = M$
that satisfy, in addition to the general  $U(1)\times U(1)$ stability subgroup conditions \p{JBpsi}, the
chirality constraints
\be
\bar{\cal D}^+\tilde{\Psi}^{(\hat N, \hat M)} =\bar{\cal D}^-\tilde{\Psi}^{(\hat N, \hat M)} = 0\,.\lb{SFgen}
\ee

\subsubsection{Supersphere superfields}

Let us consider another particular class of the general superfields $\Psi^{(\hat N, \hat M)}$
defined by \p{JBpsi}, namely those subject to the restriction
\be
{\cal D}^+\Psi^{(\hat N, \hat M)} = \bar{\cal D}^-\Psi^{(\hat N, \hat M)} = 0\, .
\lb{COND}
\ee
By virtue of the second anticommutation relation in \p{SpinCom1}, these constraints are compatible
with a non-zero superfield only when $\hat M = 0$. Then, from the definition \p{CovSF1},  it follows that
${\cal D}^+ = D^+$ and $\bar{\cal D}^- = \bar D^-$ when these operators act on the $\Psi^{(\hat N,0)}$ superfields.
Recalling the precise expressions for $D^+$ and $\bar D^-$ from  \p{CovSF2}, we conclude that \p{COND}
is equivalent, for $\hat M=0\,$, to
\be
\partial_\xi\Psi^{(\hat N, 0)} = \partial_{\bar\xi} \Psi^{(\hat N, 0)} = 0 \, .
\ee
In other words, for $\hat M=0$ the general $SU(2|1)/[U(1)\times U(1)]$ superfields may be consistently restricted,
by the covariant conditions
(\ref{COND}), to supersphere superfields, which have no dependence on the Grassmann-odd complex
coordinate $\xi$ and its complex conjugate
$\bar\xi$.

As we shall  see soon, the wave superfunctions of the quantum supersphere model indexed by $N$ belong to this subclass of
the $SU(2|1)/[U(1)\times U(1)]$ superfields in which one should identify $\hat N = N\,$.

\subsubsection{Casimir operators}

For what follows, it will be instructive to rewrite the quadratic and cubic $SU(2|1)$ Casimir operators
\p{Cas2} and \p{Cas3} in terms
of the above covariant derivatives. These are
\bea
C_2 &=& \frac{1}{2} \left(2(\hat{J}_3)^2 -\{{\cal D}^{++}, {\cal D}^{--}\} - [\bar{\cal D}^+, {\cal D}^-] -
[\bar{\cal D}^-, {\cal D}^+] \right) - (\hat{F})^2\,,   \lb{C2D} \\
C_3 &=&  \frac{1}{4}\left( \{{\cal D}^{--}, [{\cal D}^+, \bar{\cal D}^+]\} -
\{{\cal D}^{++}, [{\cal D}^-, \bar{\cal D}^-]\}\right) \nn
&& + \,\frac{1}{4}\,\{ \hat{J}_3, [{\cal D}^+, \bar{\cal D}^-]- [{\cal D}^-, \bar{\cal D}^+]\}
- \frac{1}{2}\{\hat{F}, \{{\cal D}^{++}, {\cal D}^{--}\} - 2(\hat{J}_3)^2\} \nn
&& +\, \frac{3}{4}\,\{\hat{F}, [{\cal D}^+, \bar{\cal D}^-]+[{\cal D}^-, \bar{\cal D}^+]\}
- 2 (\hat{F})^3  - \hat{F}\,. \lb{C3D}
\eea

Perhaps, the simplest way to prove the coincidence of \p{C2D} and \p{C3D} with \p{Cas2} and \p{Cas3} on
general superflag superfields is to use one more equivalent expression of the same invariant operators through
the $SU(2|1)$ generators in the manifestly $U(2)$ covariant basis:
\bea
&& C_2 =  \frac{1}{2}\,[\bar Q^k, Q_k] - F^2  - \frac{1}{2}\,T^{ik}T_{ik}\,, \lb{C2Q} \\
&& C_3 = \frac{1}{4}\,\{T^{ik}, [Q_i, \bar Q_k]\} + \frac{3}{4}\, \{F, [Q^i,\bar Q_i]\} - \frac{1}{2}\,\{F, T^{ik}T_{ik}\}
- 2 F^3 - F\,. \lb{C3Q}
\eea
In this basis, the (anti)commutation relations of the superalgebra $su(2|1)$ are
\bea
&& \{Q_i, \bar Q_k \} = \epsilon_{ik} F + T_{ik}\,, \quad  \{Q_i, Q_k \} = \{\bar Q_i, \bar Q_k \} = 0\,, \nn
&& [T_{ik}, Q_l] = \frac{1}{2}\left(\epsilon_{il} Q_k + \epsilon_{kl} Q_i\right), \quad
[T_{ik}, \bar Q_l] = \frac{1}{2}\left(\epsilon_{il} \bar Q_k + \epsilon_{kl} \bar Q_i\right), \nn
&& [F, Q_l] = \frac{1}{2}\,Q_l\,,\quad  [F, \bar Q_l] = -\frac{1}{2}\,\bar Q_l\,, \nn
&& [T_{ik}, T_{lj}] = \epsilon_{ij}T_{kl} + \epsilon_{kl}T_{ij}\,, \quad \bar Q^i =(Q_i)^\dagger\,.  \lb{su21}
\eea

\subsection{Supersphere in terms of the superflag superfields}

As was noticed in Section 2.3 (eq. \p{HC2}), the supersphere Hamiltonian \p{SSham} coincides with the Casimir operator $C_2\,$.
On the other hand, in Section 2.4, based on considering invariant norms, it was anticipated that the quantum supersphere
model at $2N$ is equivalent to the particular case of the quantum superflag model at $2N' = 2N-1$ and $M=0\,$. Following this
observation, we are led to consider the operator \p{C2D} at $\hat M = M=0\,$, i.e. at
\be
(\hat F)^2 - (\hat J_3)^2 = 0\,,
\ee
as the appropriate `would-be'  Hamiltonian of the supersphere in the manifestly covariant formulation through
superflag superfields
\be
H = -\frac{1}{2}\left({\cal D}^{++}{\cal D}^{--} + {\cal D}^{--}{\cal D}^{++} \right) + \frac{1}{2}[{\cal D}^-, \bar{\cal D}^+]
+ \frac{1}{2}[{\cal D}^+, \bar{\cal D}^-]\,. \lb{Ham1}
\ee
In Section \ref{sec:ssrevisit} we shall prove that this operator indeed reduces to \p{SSham} on the properly constrained
$SU(2|1)/[U(1)\times U(1)]$  superfields. Moreover, being restricted to the general superflag model wave functions \p{SFgen}
(with $\hat M = M\neq 0$),
it reduces to the covariant form of the superflag model Hamiltonian \p{hat2} (modulo a constant shift, see eq. \p{Hh} below).
Thus it can be regarded as a sort of `master' Hamiltonian for these two different quantum models.

Now we present the covariant form of the supersphere wave functions satisfying \p{COND}.
The supersphere LLL wave function $\Psi_0^{(N,0)}$ is defined by the conditions
\be
\mbox{(a)}: \ {\cal D}^+\Psi_0^{(N,0)} = \bar{\cal D}^-\Psi_0^{(N,0)} = 0\,; \qquad \mbox{(b)}:\
\bar{\cal D}^+\Psi_0^{(N,0)} = {\cal D}^{++}\Psi_0^{(N,0)} =0 \lb{LLL}\, .
\ee
It can be easily checked that such functions are annihilated by the operator $H$ of \p{Ham1} as a consequence of the (anti)commutation relations
\p{SpinCom0}:
\be
H\,\Psi_0^{(N,0)} = 0\, .
\ee
The conditions (\ref{LLL}a) put $\Psi_0^{(N,0)}$
on the supersphere, eliminating  the dependence on $(\xi, \bar\xi)$.  Then eqs. (\ref{LLL}b)
are the covariant chirality conditions which effectively eliminate the dependence on $\bar z$ and $\bar\zeta$;
this can be made manifest  by solving these constraints.

Already on this simplest example one can explicitly see the equivalence relation between the $M=0$ superflag
model and the supersphere
model anticipated in the previous Sections. The supersphere LLL wave function can be represented as
\be
\Psi_0^{(N,0)} = {\cal D}^+\tilde{\Psi}_0^{(N-\frac{1}{2}, 0)}\,,\lb{Rel100}
\ee
where $\tilde{\Psi}_0^{(N-\frac{1}{2}, 0)}$ obeys the constraints
\be
\bar{\cal D}^-\tilde{\Psi}_0^{(N -\frac{1}{2},0)} =
\bar{\cal D}^+\tilde{\Psi}_0^{(N -\frac{1}{2},0)} = {\cal D}^{++}\tilde{\Psi}_0^{(N-\frac{1}{2},0)} =0\,.\lb{Rel101}
\ee
The wave function defined by \p{Rel100} satisfies the constraints \p{LLL} as a consequence
of \p{Rel101} and the (anti)commutation relations \p{SpinCom0} at $M=0\,$ (in particular,
the relation ${\cal D}^+{\cal D}^+ = 0$). Now let us examine the superfunction $\tilde{\Psi}_0^{(N-\frac{1}{2}, 0)}\,$.
It is covariantly chiral and analytic. We should also take into account that
on the general set of superfunctions $\tilde{\Psi}^{(\hat N, \hat M)}$ obeying the chirality conditions \p{SFgen}
the operator \p{Ham1} is reduced, modulo a constant shift by $2\hat M$, to the superflag Hamiltonian in the covariant
formulation \cite{Ivanov:2004yw}
\be
H \quad \Rightarrow \quad H_{SF}' = -{\cal D}^{--}{\cal D}^{++} - 2\hat M = H_{SF}- 2\hat M \,.\lb{Hh}
\ee
The superfunction $\tilde{\Psi}_0^{(N-\frac{1}{2}, 0)}$ is a particular case of these general chiral functions corresponding
to $\hat N = N' = N-\frac{1}{2}\,, \hat M =M = 0$ and satisfying the additional analyticity condition
${\cal D}^{++}\tilde{\Psi}_0^{(N-\frac{1}{2}, 0)}= 0\,$. Hence, $\tilde{\Psi}_0^{(N-\frac{1}{2}, 0)}$ is just the LLL
wave function for the $N' = N-\frac{1}{2}$ superflag model at $\hat M = M=0\,$. It should be pointed out
that $\Psi_0^{(N,0)} $ is Grassmann-odd if $\tilde{\Psi}_0^{(N-\frac{1}{2}, 0)}$ is Grassmann-even and vice versa.
The first option precisely matches with our previous choice of the Grassmann parity of the wave superfunctions in the
supersphere and superflag models. Note that \p{Rel100} admits gauge invariance
\bea
&& {\tilde{\Psi}_0^{(N-\frac{1}{2}, 0)}}{\,}' = {\tilde{\Psi}_0}^{(N-\frac{1}{2}, 0)} + \Lambda^{(N-\frac{1}{2}, 0)}\,, \lb{GauGe}\\
&&{\cal D}^+\Lambda^{(N-\frac{1}{2}, 0)} = \bar{\cal D}^+\Lambda^{(N-\frac{1}{2}, 0)} =
\bar{\cal D}^-\Lambda^{(N-\frac{1}{2}, 0)} = {\cal D}^{++}\Lambda^{(N-\frac{1}{2}, 0)}=0\,. \nonumber
\eea
This can be used to remove half the component fields from $\tilde{\Psi}_0^{(N-\frac{1}{2}, 0)}$ and to equate the numbers of
the independent component fields in the left-hand and right-hand sides of \p{Rel100}; this number is just $(2 +2)$,
i.e. that of the `ultrashort' multiplet of $SU(2|1)$.

To single out, in the variety of the $SU(2|1)/[U(1)\times U(1)]$ superfields, the supersphere wave functions
related to an $\ell >0$ level, we will proceed in the following two-step way. First, we define the Grassmann-odd function
\be
\Psi_\ell^{(N,0)} = ({\cal D}^{--})^{\ell} \Phi^{(N + \ell, - \frac{\ell}{2})}\,, \lb{LL}
\ee
where the relevant `ground state wave function' $\Phi^{(N + \ell, - \frac{\ell}{2})}$ is also fermionic
and satisfies the conditions
\be
\mbox{(a)}:\ {\cal D}^+\Phi^{(N + \ell, - \frac{\ell}{2})} = 0\,;
\qquad \mbox{(b)}:\   \bar{\cal D}^+\Phi^{(N + \ell, - \frac{\ell}{2})}
= {\cal D}^{++}\Phi^{(N + \ell, - \frac{\ell}{2})} = 0\,. \lb{LLcond}
\ee
It is straightforward to check that
\be
H\,\Psi_\ell^{(N,0)} = (2N\ell + \ell^2)\,\Psi_\ell^{(N,0)}\,,  \lb{EigB}
\ee
as a consequence of \p{SpinCom0} - \p{Comm2} and \p{LLcond}. Thus,  $\Psi_\ell^{(N,0)}$
is an eigenfunction of the operator \p{Ham1} with the same eigenvalue as  in \p{casfunction}.
Note that the condition (\ref{LLcond}a) still leaves the dependence on
$\bar\xi$ in $\Phi^{(N + \ell, - \frac{\ell}{2})}\,$.

Another set of  eigenfunctions of the `would-be' Hamiltonian (\ref{Ham1}) is
\be
\widehat{\Psi}_\ell^{(N,0)} = {\cal D}^-({\cal D}^{--})^{\ell -1}
\widehat{\Phi}^{(N + \ell -\frac{1}{2}, - \frac{\ell}{2})}\,, \lb{LL2}
\ee
where
\be
\widehat{\Phi}^{(N + \ell -\frac{1}{2}, - \frac{\ell}{2})}
= \bar{\cal D}^-\Phi^{(N + \ell, - \frac{\ell}{2})}\,. \lb{Rel}
\ee
It is easy to check that
\be
H\,\widehat{\Psi}_\ell^{(N,0)} = (2N\ell + \ell^2)\,\widehat{\Psi}_\ell^{(N,0)}\,.\lb{EigF}
\ee
The bosonic reduced wave function $\widehat{\Phi}^{(N + \ell -\frac{1}{2},
- \frac{\ell}{2})}$ satisfies the conditions
\be
\mbox{(a)}:\,  \bar{\cal D}^-\widehat{\Phi}^{(N + \ell -\frac{1}{2}, - \frac{\ell}{2})} = 0\,, \qquad
\mbox{(b)}:\, \bar{\cal D}^+\widehat{\Phi}^{(N + \ell -\frac{1}{2}, - \frac{\ell}{2})}
= {\cal D}^{++} \widehat{\Phi}^{(N + \ell -\frac{1}{2}, - \frac{\ell}{2})} =0\,, \lb{LL3}
\ee
which follow from \p{LLcond} on taking into account the relations \p{SpinCom}.
Using the relation \p{Rel}, eq.\p{LL2} can be rewritten as
\be
\widehat{\Psi}_\ell^{(N,0)} = {\cal D}^-\bar{\cal D}^-({\cal D}^{--})^{\ell -1}
{\Phi}^{(N + \ell, - \frac{\ell}{2})}\,. \lb{LL4}
\ee
In other words, both series of eigenfunctions can be produced from the single fermionic
`ground state' wave function $\Phi^{(N + \ell, - \frac{\ell}{2})}$ by applying to it the operators
${\cal D}^{--}$ and ${\cal D^-}\bar{\cal D}^-\,$.  The latter operator can appear only once,
as in \p{LL4}, due to the reduction relation
\be
({\cal D^-}\bar{\cal D}^-)\,({\cal D^-}\bar{\cal D}^-) = - ({\cal D^-}\bar{\cal D}^-)\, {\cal D}^{--}\,,
\ee
which follows from \p{SpinCom0}.  Actually,
$\widehat{\Phi}^{(N + \ell -\frac{1}{2}, - \frac{\ell}{2})}$ is just the covariant definition
of the highest component in the $\bar\xi$-expansion of $\Phi^{(N + \ell, - \frac{\ell}{2})}\,$.

The next step consists in representing
the `reduced wave function' $\Phi^{(N+ \ell, -\frac{\ell}{2})}$ in \p{LL} as
\be
\Phi^{(N+ \ell, -\frac{\ell}{2})} = {\cal D}^+\tilde{\Phi}^{(N+ \ell -\frac{1}{2}, -\frac{\ell}{2})}\,. \lb{SSSF2}
\ee
The `prepotential' $\tilde{\Phi}^{(N+ \ell-\frac{1}{2}, -\frac{\ell}{2})}$ is assumed to satisfy the conditions
\be
\bar{\cal D}^-\tilde{\Phi}^{(N+ \ell-\frac{1}{2}, -\frac{\ell}{2})} =
\bar{\cal D}^+\tilde{\Phi}^{(N+ \ell-\frac{1}{2}, -\frac{\ell}{2})}
= {\cal D}^{++}\tilde{\Phi}^{(N+ \ell-\frac{1}{2}, -\frac{\ell}{2})} = 0\, ,  \lb{Rel200}
\ee
and hence can be identified with the level $\ell$ reduced wave function of the superflag model with the $U(1)$
charges $2N' = 2N - 1, M=0\,$.
The corresponding full wave functions are defined by
\be
\tilde{\Psi}^{(N-\frac{1}{2}, 0)}_\ell =
({\cal D}^{--})^{\ell}\tilde{\Phi}^{(N+ \ell-\frac{1}{2}, -\frac{\ell}{2})} \, , \lb{Rel201}
\ee
and on them the operator $H$ is reduced (cf. \p{Hh}) to
\be
H \;\Rightarrow \; H_{SF(M=0)} = -{\cal D}^{--}{\cal D}^{++}\,, \lb{HhM0}
\ee
with the eigenvalues
\be
E_\ell = (2N-1)\ell +\ell(\ell +1)\,.
\ee
The constraints (\ref{LLcond}) are satisfied as a consequence of \p{Rel200} and the relation ${\cal D}^+{\cal D}^+$ $ = 0\,$. There emerge
no additional constraints on $\Phi^{(N+ \ell, -\frac{\ell}{2})}$. Using the second relation in \p{SpinCom1}, one obtains
\be
\bar{\cal D}^-\Phi^{(N+ \ell, -\frac{\ell}{2})} = -\ell\,\tilde{\Phi}^{(N+ \ell-\frac{1}{2}, -\frac{\ell}{2})}\,.\lb{SFSS2}
\ee
For $\ell = 0$ this reproduces the first of the LLL supersphere wave function constraints (\ref{LLL}b),
while for $\ell \geq 1$ it yields the relation inverse to \p{SSSF2}. So, at $\ell \neq 0$ the relation \p{SSSF2}
is invertible (this property replaces the gauge invariance \p{GauGe}
of the LLL case).

Using \p{SSSF2} and \p{SFSS2}, as well as the (anti)commutation relations \p{SpinCom0}, one can express
both previously defined auxiliary functions \p{LL} and \p{LL2} through the $(N-\frac{1}{2}, M=0)$ superflag reduced
wave function $\tilde{\Phi}^{(N+ \ell-\frac{1}{2}, -\frac{\ell}{2})}$:
\bea
&& \Psi^{(N,0)}_\ell + \widehat{\Psi}^{(N,0)}_\ell \equiv \Psi^{(N)}_\ell =
{\cal D}^+\left[({\cal D}^{--})^{\ell}\tilde{\Phi}^{(N+ \ell-\frac{1}{2}, -\frac{\ell}{2})} \right]
= {\cal D}^+\tilde{\Psi}^{(N-\frac{1}{2}, 0)}_\ell\,, \lb{SSwf}\nn
&&\widehat{\Psi}^{(N,0)}_\ell = -\ell\,{\cal D}^-({\cal D}^{--})^{\ell-1}\tilde{\Phi}^{(N+ \ell-\frac{1}{2}, -\frac{\ell}{2})}\,. \lb{Rel300}
\eea
It is easy to show that the function $\Psi^{(N)}_\ell$ defined in \p{SSwf} satisfies both the supersphere conditions \p{COND}.
The first one
is obeyed due to the property ${\cal D}^+{\cal D}^+ = 0$, while the second one due to chirality of
$\tilde{\Phi}^{(N+ \ell-\frac{1}{2}, -\frac{\ell}{2})}\,$. This property can be made manifest using the relation \p{SFSS2}:
\be
\Psi^{(N)}_\ell  = - \frac{1}{\ell}\,({\cal D}^+\bar{\cal D}^-) \Psi^{(N,0)}_\ell =
\frac{1}{\ell}\,({\cal D}^+\bar{\cal D}^-) \widehat{\Psi}^{(N,0)}_\ell\,. \lb{Rel500}
\ee

Using the general transformation law \p{GenTranSF} of the superflag superfields $\Psi^{(\hat N, \hat M)}$
under the odd $SU(2|1)$ transformations,
one immediately observes that the wave function $\Psi^{(N,0)}_\ell$ \p{SSwf}, as well as the LLL wave function $\Psi^{(N)}_0$ defined in \p{LLL},
have the same transformation properties as the similar supersphere wave superfunctions defined in Section 2.4 (eqs. \p{tranLLL} and \p{tranLL}).
On top of this,
these constrained $SU(2|1)/[U(1)\times U(1)]$ superfields satisfy the basic condition \p{COND}, i.e. live on the supersphere, and are the
eigenfunctions of the operator $H$ \p{Ham1} with the correct eigenvalues \p{casfunction}. As shown in the Section 5.4,  this `would-be' Hamiltonian
becomes exactly \p{SSham} when applied to these superfields, which may therefore be identified with
the supersphere wave superfunctions defined in \p{LLLSS} and \p{Psi} - \p{5}; this justifies the use of the same
notation for both sets of superfields. In Section 5.4 we shall also show how the reduced superfields $\Phi^{(\pm)}_\ell$ defined in \p{3}
appear within the covariant $SU(2|1)/[U(1)\times U(1)]$ superfield approach.

Now let us discuss the relation between the $SU(2|1)$ invariant integration measures on the supersphere and superflag. In accordance
with the definitions \p{ssnorm}, \p{meas1} and \p{meas2} they are
\be
d\mu_{(SS)} = d\mu_0 (K_2)^{-1}\,, \quad d\mu_{(SF)} = d\mu_0\partial_\xi\partial_{\bar\xi}(K_2)^{-2}\,. \lb{measures}
\ee
Using (i) the relations
\be
\partial_\xi = K_2^{\frac{1}{2}}D^+\,, \quad \partial_{\bar\xi} = - K_2^{\frac{1}{2}}\bar{D}^-\,,
\ee
(ii) the fact that $K_2$ has no dependence on $\xi, \bar\xi$ and (iii) that ${\cal D}^+, \bar{\cal D}^- $ have $\hat{B}$ charge zero,
and assuming that the integrands in the corresponding integrals also have zero $\hat{B}$ charge, the measures in \p{measures}
are related by
\be
d\mu_{(SF)} = d\mu_{(SS)}\bar{\cal D}^-{\cal D}^+\,. \lb{relmeas}
\ee
Recalling the relations \p{Rel100}, \p{SSwf}  between the wave superfunctions of the supersphere and superflag models,
which can be concisely written as
\be
\Psi^{(N)}_{SS} = {\cal D}^+\Psi^{(N-\frac{1}{2}, 0)}_{SF}\,, \quad \bar{\cal D}^-\Psi^{(N-\frac{1}{2}, 0)}_{SF}
= {\cal D}^+[ \Psi^{(N-\frac{1}{2}, 0)}_{SF}]^* =0\,,
\ee
one gets the following simple relation between the inner products on the $M=0$ superflag and supersphere:
\bea
\langle \Upsilon^{(N-\frac{1}{2}, 0)}_{SF}\vert \Psi^{(N-\frac{1}{2}, 0)}_{SF}\rangle &=& \int
d\mu_{(SF)}[ \Upsilon^{(N-\frac{1}{2}, 0)}_{SF}]^*\Psi^{(N-\frac{1}{2}, 0)}_{SF} \nn
 &=& \int d\mu_{(SS)}[\Upsilon^{(N)}_{SS}]^*\Psi^{(N)}_{SS} \equiv
\langle \Upsilon^{(N)}_{SS}\vert \Psi^{(N)}_{SS}\rangle\,. \lb{inner12}
\eea
This is the superfield form of the relation between the supersphere and superflag norms observed earlier
at the component level.

It should be pointed out that the supersphere wave functions have zero norm with respect to
the superflag inner product (this directly stems from \p{relmeas} and the supersphere conditions \p{COND}) but
their norm is non-vanishing with respect to the supersphere inner product, i.e. when it is computed by the formula \p{inner12}.
Also, it is easy to check that any supersphere wave function is orthogonal to any $M=0$ superflag wave function:
their superflag inner products are vanishing. Thus the operator $H$ of  \p{Ham1} has the unique normalizable LLL
ground state with respect to the superflag inner product (recall that $H$ is reduced to
the superflag model Hamiltonian on the set of the covariantly chiral $SU(2|1)/[U(1)\times U(1)]$ superfields,
eq. \p{Hh}). The supersphere wave function $\Psi^{(N)}_0$ has zero norm,  and the possibility of adding
to it the LLL ground state $M=0$ superflag wave function  provides the gauge invariance
that is responsible for the fact that half of the component wave functions in the superflag LLL
wave superfunction at $M=0$ do not appear in its norm \cite{Ivanov:2004yw}. On the other hand, on the supersphere
wave superfunctions the same operator $H$ \p{Ham1} is reduced to the supersphere Hamiltonian \p{SSham} (see Section 5.4),
with the {\it same} $\Psi^{(N)}_0$ as the LLL wave function. The latter has a non-zero norm with respect to the inner product
on the supersphere.

To summarize, at  given fixed $\hat N=N$, the $M=0$ superflag model wave functions and the supersphere
model wave functions span two different subspaces, closed under the action of $SU(2|1)$,
in the full variety of  $SU(2|1)/[U(1)\times U(1)]$ superfields. These subspaces are
orthogonal to each other with respect to the natural inner product on $SU(2|1)/[U(1)\times U(1)]\,$.
The supersphere wave functions have zero norm with respect to this product, but non-vanishing norm
with respect to the inner product on the invariant submanifold $SU(2|1)/U(1|1) \subset SU(2|1)/[U(1)\times U(1)]$.
The operator $H$ is independently diagonalized on each of these two mutually orthogonal subspaces and
is reduced on them, respectively, to the supersphere Hamiltonian \p{SSham} and to the
$M=0$ superflag Hamiltonian \p{HhM0}. Taken at the {\it same fixed} $\hat N = N$, these two models
are not equivalent to each other.
The $N$ supersphere model is equivalent to the $M=0$ superflag model with $N' = N-\frac{1}{2}$,
and the covariant formulation given in this Section makes this equivalence manifest.

\subsection{Casimir considerations}

To better understand the difference between the wave functions of the quantum supersphere and superflag models
in the manifestly covariant
unified description, let us compare the values of the Casimir operators \p{C2D} and \p{C3D} on these wave functions.
The subsequent consideration is based on the fact that all covariant derivatives defined in \p{CovSF1}, \p{CovSF2}
{\it commute} with
the Casimir operators $C_2$ and $C_3\,$, so the values of the latter can be evaluated by applying them directly
to the reduced wave functions.

For the general $M=0$ superflag `ground state' wave functions $\tilde{\Phi}^{(\tilde N, \tilde M)}$ subjected
to the covariant chirality and analyticity conditions
\be
\bar{\cal D}^+\tilde{\Phi}^{(\tilde N, \tilde M)} = \bar{\cal D}^-\tilde{\Phi}^{(\tilde N, \tilde M)} =
{\cal D}^{++}\tilde{\Phi}^{(\tilde N, \tilde M)}
 = 0\, ,  \lb{SFconstr}
\ee
one finds
\bea
C_2 = -2\tilde M [1 + 2(\tilde M + \tilde N)]\,, \nonumber
\eea
\bea
C_3 = -2\tilde M[1 + 2(\tilde M + \tilde N)] [1 + 2(\tilde N + 2\tilde M)] = [1 + 2(\tilde N + 2\tilde M)]C_2\,. \lb{CSF1}
\eea
Both Casimirs vanish on the LLL state with $\tilde N= N, \tilde M =M =0$, which corresponds to the `atypical'
representation of $SU(2|1)$. For the $\ell$-th LL `ground state' with
$\tilde N = N + \ell \,, \tilde M = -\frac{\ell}{2}$ the Casimirs take the values
\be
C_2 = \ell (1 + \ell + 2N)\,, \quad C_3 = \ell(1 + 2N)(1 + \ell + 2N)\,. \lb{CSF2}
\ee

For the general supersphere bosonic `ground state' wave functions $\Phi^{(\tilde N, \tilde M)}$
subjected to another sort of  Grassmann analyticity conditions, and the standard bosonic analyticity condition
\be
{\cal D}^+{\Phi}^{(\tilde N, \tilde M)} = \bar{\cal D}^+{\Phi}^{(\tilde N, \tilde M)} = {\cal D}^{++}{\Phi}^{(\tilde N, \tilde M)} = 0\,,
\ee
one finds that
\be
C_2 = -4\tilde M (\tilde M + \tilde N)\,, \quad C_3 = - 8\tilde M(\tilde N + 2\tilde M)(\tilde N + \tilde M)
= 2(\tilde N + 2\tilde M)C_2\,. \lb{CSS1}
\ee
On the supersphere LLL state with $\tilde N = N, \tilde M =0$, these Casimirs again vanish, showing that
this $SU(2|1)$ multiplet is also `atypical'.
However, for any other LL with
$\tilde N = N + \ell \,, \tilde M = -\frac{\ell}{2}$ the Casimirs
take the values
\be
C_2 = 2\ell \left(N + \frac{\ell}{2}\right), \quad C_3 = 4 N\ell \left(N + \frac{\ell}{2}\right),  \lb{CSS2}
\ee
which does not coincide with \p{CSF2} at the {\it same} fixed $N$. Thus in the superflag and supersphere cases at the same {\it fixed}
$N$ we deal with {\it different}
representations of the supergroup $SU(2|1)$.
Comparing \p{CSF2} and \p{CSS1} one observes that the $M=0$ superflag wave function backgrounds
can be obtained from the supersphere ones by the substitution $N \rightarrow N + \frac{1}{2}$ in the latter.
This correspondence just amounts to the equivalence of the $N$ supersphere model and the $N' = N-\frac{1}{2}, M=0$
superflag model, as established in the previous Sections.

Finally, let us establish the precise relation between the operator $C_2$ \p{C2D} and the superflag
model Hamiltonian in the covariant formulation, i.e. with
\be
H_{SF} = -{\cal D}^{--}{\cal D}^{++}\,.
\ee
Using the properties that on the general superflag model superfields \p{SFgen}
\be
C_2 = H + (\hat J_3)^2 - (\hat F)^2 = H + (N')^2 - (N' + 2M)^2\,,
\ee
and, according to \p{Hh},
\be
H = H_{SF} - 2M\,,
\ee
we find that $C_2 = H_{SF} - 2M\left(2M + 2N' + 1\right)$, in full agreement with eq. \p{HSF11}.

\subsection{The supersphere model revisited}
\label{sec:ssrevisit}

As the last topic of this Section we establish the explicit relation with the consideration of Section 2.
Here we make use of the notation $P$ to denote a point on the `supersphere' superspace $(z,\bar z,\zeta,\bar\zeta)$.

After a rather tedious calculation, the Hamiltonian operator \p{Ham1}, being applied to a general wave functions $\Psi^{(N,0)}$
defined by \p{JBpsi} with $N' =N, M' =0$, can be cast in the following explicit form
\bea
&& H = H_0 -\frac{\partial}{\partial \xi }\, \frac{\partial}{\partial \widehat{\bar{\xi_{\,}}}}
- K_2\left(\bar\zeta + \bar z  \widehat{\bar{\xi_{\,}}}\right)
\left(\nabla_{\bar z}^{(N)} - \bar\xi\nabla_{\bar \zeta}^{(N)}\right)\frac{\partial}{\partial \widehat{\bar{\xi_{\,}}}} \nn
&& +\, K_2^{-1} \bar z\left[\left(1 + z\bar z\right)\xi + \bar z \zeta \right]\left(\nabla_{\bar z}^{(N)}
- \bar\xi \nabla_{\bar \zeta}^{(N)}\right)\frac{\partial}{\partial \xi } \nn
&& - \,\left\{\bar z\left[K_1 - (\zeta\bar\zeta)(\xi\bar\xi)\right]
- \xi \bar\zeta \right\}\nabla_{\bar \zeta}^{(N)}\,\frac{\partial}{\partial \xi }\,. \lb{HamPrec}
\eea
Here $\widehat{\bar{\xi_{\,}}}$ was already defined in \p{Ext1}
\be
\widehat{\bar{\xi_{\,}}} = \bar\xi\,K_2 + \bar\zeta\,z\,, \lb{hat}
\ee
and $\bar\xi$ is assumed to be expressed through $\widehat{\bar{\xi_{\,}}}$ from \p{hat},
\be
\bar\xi = K_2^{-1}\left(\widehat{\bar{\xi_{\,}}} - z\bar\zeta \right).\lb{hat-}
\ee
The part $H_0$
coincides with the Hamiltonian \p{SSham}:
\be
H_0 = -g^{z\bar z}\nabla_{z}^{(N)}\nabla_{\bar z}^{(N)} - g^{\zeta\bar\zeta}\nabla_{\zeta}^{(N)}\nabla_{\bar \zeta}^{(N)}
+ g^{\zeta\bar z}\nabla_{\zeta}^{(N)}\nabla_{\bar z}^{(N)} - g^{z \bar\zeta}\nabla_{z}^{(N)}\nabla_{\bar\zeta}^{(N)}\,,
\ee
where the derivatives $\nabla^{(N)}_B, \nabla^{(N)}_{\bar B}$ were defined in \p{nablas}, \p{explNab}.
The Hamiltonian $H_0$ contains no derivatives with respect to the Grassmann variables $\xi, \widehat{\bar{\xi_{\,}}}$
which complement the supersphere to the superflag, so it is defined on the supersphere.

In terms of the covariant derivatives, the transition to
the variable $\widehat{\bar{\xi_{\,}}}$ of  \p{hat} eliminates
the partial derivative with respect to $\bar\xi$ from
the covariant derivative $\bar{\cal D}^+$. For what follows, it is
instructive to give the expressions for the covariant derivatives
$\bar{\cal D}^+, {\cal D}^+, \bar{\cal D}^-, {\cal D}^{\pm\pm}$ in the
new basis and in application to the superfields with $\hat{J}_3 =
2N$ and $\hat{B} =0$:
\bea
&& \bar{\cal D}^+ =
-(K_1K_2)^{\frac{1}{2}}\left\{\nabla^{(N)}_{\bar\zeta} -
K_1^{-1}\left[(1 + z\bar z)\xi + \bar z
\zeta\right]\left[\nabla^{(N)}_{\bar z} -
\bar\xi\nabla^{(N)}_{\bar\zeta} \right]\right\},\nn
&& {\cal
D}^+ = K_2^{-\frac{1}{2}}\partial_{\xi}\,,  \quad \bar{\cal D}^- =
-K_2^{\frac{1}{2}}\partial_{\widehat{\bar{\xi_{\,}}}}\,,  \nn
&&
{\cal D}^{++} = K_1^{\frac{1}{2}} K_2\left[\nabla^{(N)}_{\bar z}
-\bar\xi\nabla^{(N)}_{\bar\zeta} \right], \nn
&&{\cal D}^{--} = K_1^{\frac{1}{2}} K_2\left\{ \left[\nabla^{(N)}_{z}
-\xi\nabla^{(N)}_{\zeta} \right] + K^{-1}_2
\left[(1 + \xi\widehat{\bar{\xi_{\,}}})\bar\zeta
+ \bar z \widehat{\bar{\xi_{\,}}}\right]\partial_{\widehat{\bar{\xi_{\,}}}}\right\}.\lb{CovSF3}
\eea
Here $\bar\xi$ is assumed to be expressed as in \p{hat-}.

Now, using these explicit expressions, one can show that
the constraints \p{LLL} defining the LLL wave function $\Psi_0^{(N,0)}$ amount to the following explicit set of equations:
\bea
&& \mbox{(\ref{LLL}a)}: \quad \frac{\partial}{\partial \xi }\Psi_0^{(N,0)}
= \frac{\partial}{\partial \widehat{\bar{\xi_{\,}}}}\Psi_0^{(N,0)} = 0 \;\; \Rightarrow \;\;\Psi_0^{(N,0)} =
\Psi_0^{(N,0)}(P)\,, \nn
&& \mbox{(\ref{LLL}b)}: \quad \nabla_{\bar z}^{(N)}\Psi_0^{(N,0)} = \nabla_{\bar \zeta}^{(N)}\Psi_0^{(N,0)} = 0\,.
\eea
Thus the LLL wave function in the `superflag-inspired' formalism coincides with the LLL wave function $\Psi^{(N)}_0(P)$
defined by the constraints \p{LLLSS} of Section 2: $\Psi^{(N, 0)}_0 = \Psi^{(N)}_0$. The $SU(2|1)$ transformation of
$\Psi^{(N,0)}_0$ obtained by the general formula \p{GenTranSF} coincides with the transformation law \p{tranLLL}.
The `would-be' Hamiltonian \p{HamPrec} is reduced to the supersphere Hamiltonian $H_0$ on $\Psi^{(N,0)}_0\,$.

For the `ground state' wave function $\Phi^{(N + \ell, - \frac{\ell}{2})}$ corresponding
to the $\ell$-th LL and satisfying the conditions \p{LLcond}, one is led to make the following redefinition
\be
\Phi^{(N + \ell, - \frac{\ell}{2})} = \left(1 - \xi\widehat{\bar{\xi_{\,}}}\right)^{\ell}K_1^{-\frac{\ell}{2}} K_2^{-\ell}
\Phi_{(\ell)}\,. \lb{Redef8}
\ee
The constraint (\ref{LLcond}a) then implies
\be\lb{ExpP}
\frac{\partial}{\partial \xi }\Phi_{(\ell)} = 0\;\; \Rightarrow
\Phi_{(\ell)}(z,\bar z,\zeta,\bar\zeta, \widehat{\bar{\xi_{\,}}})
= \omega_{(\ell)}(z,\bar z,\zeta,\bar\zeta) +
\widehat{\bar{\xi_{\,}}} \phi_{(\ell)}(z,\bar z,\zeta,\bar\zeta)\, ,
\ee
while (\ref{LLcond}b) implies
\bea
&&\nabla_{\bar z}^{(N)}\Phi_{(\ell)} =
\nabla_{\bar \zeta}^{(N)}\Phi_{(\ell)} = 0 \;\; \Rightarrow \;\;\nn
&& \nabla_{\bar z}^{(N)}\phi_{(\ell)} =
\nabla_{\bar \zeta}^{(N)}\phi_{(\ell)} = 0\,, \;\; \nabla_{\bar z}^{(N)}\omega_{(\ell)} =
\nabla_{\bar \zeta}^{(N)}\omega_{(\ell)} = 0\,.
\eea

The covariantly chiral bosonic and fermionic functions $\phi_{(\ell)}$ and $\omega_{(\ell)}$
can be identified with the functions $\Phi^{(-)}_\ell$ and $\Phi^{(+)}_\ell$ defined by \p{1} - \p{3}.
Indeed, let us consider the transformation law of \p{Redef8} under the odd $SU(2|1)$ transformations.
The left-hand side of \p{Redef8} is transformed according to the general law \p{GenTranSF},
with $\hat N = N+\ell, \hat M = - \frac{\ell}{2}\,$.
Then, using the coordinate transformations \p{superstr}, \p{xiTran} and \p{barxiTran},
it is straightforward to find the transformation law of
$\Phi_{(\ell)}(z,\bar z,\zeta,\bar\zeta,\widehat{\bar{\xi_{\,}}})$:
\be
\delta \Phi_{(\ell)} = -\left[ N\left(\epsilon^1\bar\zeta +\bar\epsilon_1\zeta \right) +
\ell \left(\bar\epsilon_1\zeta - \epsilon^2\widehat{\bar{\xi_{\,}}}\right)\right]\Phi_{(\ell)}. \lb{Phiell}
\ee
Recalling the transformation law of $\widehat{\bar{\xi_{\,}}}\,$, eq. \p{barxiTran},
\be
\delta\widehat{\bar{\xi_{\,}}} = (\bar\epsilon_2 + z\bar\epsilon_1) - (\bar\epsilon_1\zeta)\widehat{\bar{\xi_{\,}}}\,,
\ee
and identifying
\be
\omega_\ell = \Phi^{(+)}_\ell\,, \quad \phi_\ell = -\ell\,\Phi^{(-)}_\ell\,,\lb{identif}
\ee
for the variations of the so defined $\Phi^{(\pm)}_\ell$ we obtain from \p{Phiell} just the expressions \p{+-transf}.

As for the non-reduced $H$-eigenfunctions $\Psi^{(N,0)}_\ell$, $\widehat{\Psi}^{(N,0)}_\ell$ related
to the `ground state' ones by eqs. \p{LL}, \p{LL3}, their relation to the functions
$\Psi^{(N)}_{(+)\ell}(z,\bar z,\zeta,\bar\zeta)$, $\Psi^{(N)}_{(-)\ell}(z,\bar z,\zeta,\bar\zeta)$
used in Section 2 is rather non-direct. We show this relation for the simplest $\ell = 1$ case.
The detailed form of the
relation between the wave functions $\Psi^{(N,0)}_1$ and $\Phi^{(N + 1, - \frac{1}{2})}$ is as follows
\bea
\Psi^{(N,0)}_{\ell =1} = {\cal D}^{--}\Phi^{(N + 1, - \frac{1}{2})} &=& \left(1 -\xi\widehat{\bar{\xi_{\,}}}\right)
\left[\nabla_{z}^{(N +1)} - \xi\nabla_{\zeta}^{(N +1)}\right]\Phi_{(1)} \nn
&& +\, K_2^{-1}\left(\bar\zeta + \bar z\widehat{\bar{\xi_{\,}}}\right )
\frac{\partial}{\partial \widehat{\bar{\xi_{\,}}}}\Phi_{(1)}
- \bar z\, K^{-1}_2\,\Phi_{(1)} \,, \lb{11}
\eea
where we made use of \p{Redef8} for $\ell = 1\,$. To calculate $\widehat{\Psi}^{(N,0)}_1$, we make use of
the explicit expression for the covariant bosonic function
$\widehat{\Phi}^{(N + \ell -\frac{1}{2}, - \frac{\ell}{2})}$ defined in \p{LL2}, \p{LL3} and related to
$\Phi^{(N + \ell, - \frac{\ell}{2})}$ by \p{Rel}. It reads
\bea
\widehat{\Phi}^{(N + \ell -\frac{1}{2}, - \frac{\ell}{2})} &=&\bar{\cal D}^-\Phi^{(N + \ell, - \frac{\ell}{2})}
\nonumber\\
&=&  -K_1^{-\frac{\ell}{2}}K_2^{- \ell +\frac{1}{2}}
\frac{\partial}{\partial \widehat{\bar{\xi_{\,}}}}\left[\left(1 - \xi\widehat{\bar{\xi_{\,}}}\right)^\ell
\Phi_{(\ell)}(z,\bar z,\zeta,\bar\zeta, \widehat{\bar{\xi_{\,}}})\right].
\eea
Then we find
\bea
\widehat{\Psi}^{(N,0)}_{\ell =1} = {\cal D}^-\bar{\cal D}^-\Phi^{(N + 1, - \frac{1}{2})}&=&
-\left[\nabla^{(N+1)}_\zeta + \widehat{\bar{\xi_{\,}}}\nabla^{(N+1)}_z\right]\left(\xi + \partial_{\widehat{\bar{\xi_{\,}}}}\right)\Phi_{(1)}
\nn
&& -\, K_2^{-1}\left(\bar\zeta + \bar z\widehat{\bar{\xi_{\,}}}\right )
\frac{\partial}{\partial \widehat{\bar{\xi_{\,}}}}\Phi_{(1)}
+ \bar z\, K^{-1}_2\,\Phi_{(1)} \,. \lb{15}
\eea
It is easy to check that, in the full agreement with the relations \p{SSwf} and \p{Rel500},
\bea
&& \Psi^{(N,0)}_{\ell =1} + \widehat{\Psi}^{(N,0)}_{\ell =1} = \nabla^{(N+1)}_z\Phi^{(+)}_1
+ \nabla^{(N+1)}_\zeta\Phi^{(-)}_1 = \Psi^{(N)}_1\,, \nn
&& \Psi^{(N)}_{1} = -({\cal D}^+\bar{\cal D}^-)\Psi^{(N,0)}_{\ell =1}
= \partial_\xi\partial_{\widehat{\bar{\xi_{\,}}}}\Psi^{(N,0)}_{\ell =1}\,.
\eea

In a similar way, using e.g. eq. \p{Rel500}, one can find that $\Psi^{(N)}_\ell =\Psi^{(N, 0)}_\ell + \widehat{\Psi}^{(N,0)}_\ell$ are
expressed through $\Phi^{(\pm)}_\ell$ just according to \p{Psi}, \p{5}. Because $\Psi^{(N)}_\ell$ satisfy the conditions $\p{COND}$, they
do not depend on $\xi, {\widehat{\bar{\xi_{\,}}}}$ and on them the operator \p{HamPrec} is reduced to $H_0$,
i.e. to the supersphere Hamiltonian.

Note that the eigenvalue relation for $\ell =1$
\be
H\,\Psi_{\ell=1}^{(N,0)} = (2N + 1)\,\Psi_{\ell=1}^{(N,0)}\,, \lb{EigB1}
\ee
can be shown to imply the relation
\bea
&& H_0 \left[\nabla_{z}^{(N +1)} -
\xi\nabla_{\zeta}^{(N +1)}\right]\Phi_{(1)}(z,\bar z,\zeta,\bar\zeta,\widehat{\bar{\xi_{\,}}}) \nn
&& = (2N + 1)\left[\nabla_{z}^{(N +1)} - \xi\nabla_{\zeta}^{(N +1)}\right]
\Phi_{(1)}(z,\bar z,\zeta,\bar\zeta,\widehat{\bar{\xi_{\,}}})\,.\lb{EigB12}
\eea
Since $H_0$ contains no any derivatives with respect to $\xi,\widehat{\bar{\xi_{\,}}}$, \p{EigB12}
amounts to
\bea
&& H_0\, \nabla_{z}^{(N +1)}\Phi^{(\pm)}_{\ell =1} = (2N + 1)\,\nabla_{z}^{(N +1)}\Phi^{(\pm)}_{\ell =1}\,, \nn
&& H_0\, \nabla_{\zeta}^{(N +1)}\Phi^{(\pm)}_{\ell =1} = (2N + 1)\,\nabla_{\zeta}^{(N +1)}\Phi^{(\pm)}_{\ell =1} \,,\lb{EigB13}
\eea
where the functions $\Phi^{(\pm)}_{\ell}(z,\bar z,\zeta,\bar\zeta)$ are defined by the $\widehat{\bar{\xi}_{\,}}$ expansion in
\p{ExpP} and by \p{identif}. Analogous relations can be obtained for $\ell >1\,$.
Such eigenfunctions have complicated $SU(2|1)$ transformation laws and presumably correspond to some composite higher
superspin $SU(2|1)$ multiplets.

\subsection{Digression: $SU(2|1)/U(2)$ superfields}
For completeness, we comment here on another subclass of general
superflag superfields: those that are defined on  the purely
fermionic coset space $SU(2|1)/U(2)$. These superfields do not
depend on the coordinates $z$ and $\bar z$, so they are defined by
the following $SU(2|1)$ covariant condition
\be
{\cal
D}^{++}\Phi^{(0,\hat{M})} = {\cal D}^{--}\Phi^{(0,\hat{M})} = 0\,.
\lb{ChirII}
\ee
By virtue of the last commutation relation in
\p{Comm2}, these constraints are compatible only when $\hat{J}_3 =
\hat{N} = 0\,$, in which  case it is convenient to pass back to
the coordinates $\xi^1, \xi^2$ defined in \p{NewCo}, and used in
\cite{Ivanov:2004yw}. In these coordinates, ${\cal D}^{\pm\pm}$
involve only the partial derivatives $\partial_{z}$ and
$\partial_{\bar z}$, so it becomes manifest that the constraints
\p{ChirII} eliminate $z, \bar z$ dependence; i.e.
\be
\mbox{\p{ChirII}}\quad \Rightarrow \quad \Phi^{(0,\hat{M})} =
\Phi^{(0,\hat{M})}(\xi^1, \xi^2, \bar\xi_1,  \bar\xi_2)\,.
\ee
It is consistent to further impose on these general $SU(2|1)/U(2)$
superfields either the covariant chirality conditions
\be
\bar{\cal D}^{+}\Phi^{(0,\hat{M})}_{(1)} = \bar{\cal
D}^{-}\Phi^{(0,\hat{M})}_{(1)} = 0\,, \ee or the covariant
anti-chirality conditions \be {\cal D}^{+}\Phi^{(0,\hat{M})}_{(2)}
= {\cal D}^{-}\Phi^{(0,\hat{M})}_{(2)} = 0\,.
\ee
These
(anti)chirality constraints can be solved explicitly in terms of
`small-(anti)analytic'  superfields, $\varphi_{(1)}$ or
$\varphi_{(2)}$, which depend (anti)holomorphically on half of the
fermionic coordinates; e.g.
\be
\Phi^{(0,\hat{M})}_{(1)} =
(K_1)^{\hat M}\varphi^{(0, \hat M)}_{(1)}(\xi^1, \xi^2)\, , \qquad
K_1 = 1 - \xi^1 \bar \xi_1 -\xi^2 \bar\xi_2\, .
\ee

Just this kind of superfield appeared as a wave superfunction  in the model
of odd $SU(2|1)$ invariant quantum mechanics considered in \cite{Ivanov:2003ax}.
There,  the Lagrangian  was taken to be the fermionic WZ term corresponding
to $U(1) \subset U(2)$, so the Hamiltonian is zero and all states are described
by a  single chiral LLL wave superfunction. One could extend this model by adding
to the WZ term the square of purely fermionic coset Cartan forms. In this case
one should expect to have to consider higher Landau levels and the possibility
of ghosts.  The planar limit of such a model was studied in  \cite{Curtright:2006aq}
under the rubric `fermionic Landau model'; it was found that there are just
two Landau levels in this limit, and that ghosts can be eliminated
by an appropriate  non-trivial choice of Hilbert space metric.  We shall not pursue
this investigation further here since the $SU(2|1)/U(2)$ Landau models
cannot be considered as `spherical' super-Landau models.

%%%%%%%%%%%%%%%%%%%%%%%%%%%%
\section{Conclusions}
\setcounter{equation}{0}

This paper concludes a series of earlier  investigations into $SU(2|1)$-invariant extensions of the
$SU(2)$-invariant spherical Landau models, parametrized by an integer electric charge $2N$.
At the classical level, these models involve  additional anti-commuting variables which, upon quantization,
lead to additional quantum states in each Landau level such that  each level furnishes a representation
of the supergroup $SU(2|1)$.

The series began with a study of  the lowest Landau level  for a particle on the supersphere
$\bC\bb{P}^{(1|1)}\cong SU(2|1)/U(1|1)$, as a special case of  $\bC\bb{P}^{(n|m)}$. One may take a limit in which
only the lowest Landau level survives and in this limit the model provides a `quantum superspace'
description of the fuzzy superspheres of fuzzy degree $2N$ \cite{Ivanov:2003qq}.  The quantum states
of the lowest Landau level all have positive norm with respect to an $SU(2|1)$-invariant
inner product that is naturally defined as a superspace integral, but this inner product
implies the existence of negative norm states, or `ghosts',  in all higher levels.
This unsatisfactory state of affairs is ameliorated in the `superflag' Landau models
which are based on the coset superspace $SU(2|1)/[U(1)\times U(1)]$ and which involve
an additional  anti-commuting  variable; these models  also have an additional parameter,
$M$, which has no effect on the energy levels but does have an effect on the norms
of states \cite{Ivanov:2004yw}. For positive $M$ it was found that the first $[2M]+1$
Landau levels are ghost-free,  in the natural superspace norm, although there
are still ghosts in higher Landau levels, and in all levels for $M<0$.

An unusual feature of the superflag Landau models is that
zero-norm states appear for non-negative integer $2M$. This is due
to the existence, for non-negative $M$, of a fermionic gauge
symmetry of the classical theory within the phase-space `shell' of
energy $2M$, which has an effect on the quantum theory when $2M$
is a non-negative integer. This unusual feature was  investigated
in detail in the context of the planar limit, which yields the
`planar superflag' Landau models \cite{Ivanov:2005vh}; in
particular, it was shown that zero-norm states in the lowest
Landau level of the $M=0$ planar superflag Landau model ensure the
equivalence of this model with the `superplane' Landau model,
obtained as the planar limit  of the superspherical Landau model.
The latter is very similar to a model studied earlier by Hasebe
\cite{Hasebe:2005cm}, but  differs in the reality conditions
imposed on the anti-commuting variables.

One surprising aspect of the superplane Landau model is that the
energy spectrum is precisely that of a model of supersymmetric
quantum mechanics, at least if one quantizes in such a way that
the state space is a conventional Hilbert space and not a vector
superspace. This feature implies the existence of an alternative
positive norm, with respect to which the superplane  Landau model
is both unitary and `worldline' supersymmetric (and this is
implicit in Hasebe's  work on his `superplane' model) but it is
not obvious that a positive norm will preserve the original
`internal' supersymmetry that motivated the model's construction.
The planar super-Landau models  were `revisited'  in
\cite{Curtright:2006aq} with the aim of clarifying this point.
It was found that the `internal' supersymmetry permits two
possible norms, such that  the alternative norm is positive  when
$M\le 0\,$; a `dynamical' combination of the two norms is needed for
positivity when $M>0\,$. A redefinition of the norm also changes the
definition of hermitian conjugation, such that the new hermitian
conjugates are `shifted' by operators that generate `hidden'
symmetries. Remarkably, the non-zero  `shift' operators were
found, for $M\le0$, to be the odd generators of a hidden worldline
supersymmetry, spontaneously broken for $M<0$ but unbroken for
$M=0\,$.

In this paper we have carried out a similar analysis for the
superspherical Landau model, and for the associated superflag
Landau models. One result of our analysis is the proof of a
quantum equivalence between the $M=0$ superflag Landau model with charge
$2N'= 2N-1$ and the superspherical Landau model with charge $2N$.
Classically,  there is an equivalence between these models for the {\it same} charge
provided the energy is non-zero, so the `quantum shift' of the charge
by one unit is presumably due to some effect associated with zero energy
configurations.

We have shown that $SU(2|1)$ invariance of the general superflag model
allows a positive Hilbert space norm that is a
`dynamical' combination of the `naive' superspace norm and an
`alternative' norm that involves a non-trivial Hilbert space `metric operator'.
This alternative norm leads, by itself, to  a unitary model when $-2N<2M\le 0$, and
these are the cases that we have focused on.  We have `solved'  these
unitary models for all $N$: that is to say,  we have found the
complete $SU(2|1)$ representation content at each Landau level. If
it had been appreciated from the outset  that $SU(2|1)$ invariance
is compatible with unitarity then it is possible that the
superspherical Landau models would have been solved  directly
without the detour into the superflag Landau models, but  the
detour has proved instructive; the superflag models are simpler in
some respects, and the superspherical models can be obtained by restricting
to  $M=0$.

An interesting general issue, not investigated here,  is  how the semi-classical limit is modified
by a  change in the Hilbert space metric. In the coherent state  approach to the
classical limit,  the symplectic 2-form associated to the classical dynamics clearly
depends on the Hilbert space metric. A change from a non-positive metric to a positive
one cannot be unitary, so we should expect a non-canonical transformation of the
classical  phase space. However, the negative norms that we find for the `naive'
Hilbert space metric are associated with the anti-commuting variables for which
there is no truly classical limit, but this  issue may be of interest in the context of the quasi-
hermitian \cite{Scholtz:1992jg, Bender:2005tb} quantum mechanics,
where it is well known that the non trivial Hilbert space metric plays
a central role.

One of our objectives in this paper was to see whether the hidden
worldline supersymmetry of the unitary planar super-Landau models
is inherited from some analogous symmetry of unitary spherical
super-Landau models. The introduction of a non-trivial `metric
operator', required to relate the alternative norm to the `naive'
one, implies the redefinition of some hermitian conjugates by
`shift' operators that are guaranteed by the formalism to be new
`hidden' symmetry generators. There is no guarantee that such
`hidden' symmetries will close to yield a finite-dimensional
enlarged symmetry algebra but  a closed subset can be found for
 $-2N < 2M <0$. In these cases the manifest $SU(2|1)$ symmetry is a
subgroup of an $SU(2|2)$ symmetry\footnote{The supergroup $SU(2|2)$
also arises in the context of integrable spin chains of relevance
to the planar limit of  ${\cal N}=4$ super-Yang-Mills theory \cite{Beisert:2006qh},
but we are not aware of any connection to our work. More recently,
$SU(2|2)$ (actually, its some non-linear version) was identified as a `hidden' symmetry  of a model
of ${\cal N}=2$  supersymmetric Quantum Mechanics \cite{Correa:2007fi};
again, we are not aware of any relation to our work.}with a central charge that is
linear in the `level operator'. The $M=0$ case is similar in many
respects but the lowest Landau level is now special and this
prevents any simple construction of  a finite  basis of  charges
with level-independent (anti)commutation relations; it thus seems
likely that  any symmetry group of the superspherical Landau model
that contains $SU(2|1)$ but has higher dimension will have
infinite dimension\footnote{This conclusion may be contrasted with
claims made for the alternative $OSp(1|2)$-invariant
`superspherical'  Landau model studied in \cite{Hasebe:2004hy},
but any disagreement could be a consequence of a quantum
inequivalence to the  superspherical Landau model considered
here.}.

Finally, one may hope that the unitary super-Landau models analyzed
here and in our previous papers will find applications. One possibility
is that they may provide an improved framework for the recently proposed
\cite{ST2008} Landau-model approach to the Riemann hypothesis.

\section*{Acknowledgements}
We benefited from useful conversations and remarks from C.
Bunster, R. Casalbuoni, J. Gauntlett, G. Gibbons,  Jaume Gomis ,
Joaquim Gomis, M. Henneaux, V. Rittenberg,  A. Schwimmer, L.
Susskind,  G. Veneziano,  and A. B. Zamolodchikov.  We also
acknowledge helpful comments of an anonymous referee. T.C. and
L.M. acknowledge partial support from the National Science
Foundation under grant No 0555603. L.M. acknowledges partial
supports from the University of Miami Provost's Award for
Scholarship Activity, from  the organizers of the Galileo Galilei
Institute Workshop ``String and M Theory approaches to particle
physics and cosmology"  and from JINR (Dubna, Russia). Parts of
these investigations were presented by L.M. in the conference
talks at Galileo Galilei Institute Workshop ``String and M Theory
approaches to particle physics and cosmology'', ``Supersymmetries
and Quantum Symmetries",  July 30 - August 4, 2007, Dubna, Russia
and ``Claudio's Fest: QUANTUM MECHANICS OF FUNDAMENTAL SYSTEMS:
THE QUEST FOR BEAUTY AND SIMPLICITY'', January 10-11, 2008,
Valdivia, Chile. E.I. acknowledges a partial support from RFBR
grants, projects  No.  06-02-16684-a and No. 08-02-90490, the
grant INTAS-1000005-7928, the DFG grant No.436 RUS 113/669/0-3,
and a grant of the Heisenberg-Landau program. P.K.T. thanks the
EPSRC for financial support.

%%%%%%%%%%%%%%%%%%%%%%%%%%%%%
\section*{ Appendix: $C\! P^n$ Landau model}
\setcounter{equation}{0}
\renewcommand{\theequation}{A-\arabic{equation}}
In this appendix we show how the method used in Section \ref{sect:spectrum}
to solve the supersphere
model can be applied to the Landau model for a particle on $\bC\bb{P}^n$,
which we view  as a K\"ahler manifold of complex dimension $n$
with isometry group $SU( n+1)$. One may choose complex coordinates
$\{z^a; a=1,\dots,n\}$ such that the  K\"ahler potential is
\be
{\cal K}   = \log\left(1 + \bar z \cdot z\right)   \, , \qquad
\left(\bar z \cdot z = \sum_{a=1}^n \bar z^a z^a\right).
\ee
The corresponding K\"ahler metric is
\be\label{Kmetric}
g_{\bar b a} \equiv \partial_{\bar b} \partial_a  {\cal K} =
\left(1+ \bar z \cdot z\right)^{-1} \left[ \delta_{ab}
- \left(1+ \bar z z\right)^{-1} z^b \bar z^a\right]\, ,
\ee
where we use the notation
\be
\partial_a = \frac{\partial}{\partial z^a}\, , \qquad
\partial_{\bar a} = \frac{\partial}{\partial \bar z^a}\, .
\ee The K\"ahler potential ${\cal A}$ for the K\"ahler 2-form
${\cal F}=d{\cal A}$ is \be {\cal A} \equiv - i \left(dz^a
\partial_a   - d\bar z^b \partial_{\bar b}\right){\cal K} =  dz^a
{\cal A}_a   + d\bar z^b {\cal A}_{\bar b}\, , \ee which gives \be
{\cal A}_a = -i \frac{\bar z^a}{1+ \bar z z} \, , \qquad {\cal
A}_{\bar b} = i \frac{z^b}{1+\bar z z}\, .
\ee
With these
ingredients we may write down the classical Lagrangian for the
$\bC\bb{P}^n$ Landau model:
\be\label{CPlag}
L= \dot z^a \dot{\bar z}^b
g_{\bar b a} + N\left(\dot z^a {\cal A}_a + \dot{\bar z}^b {\cal
A}_{\bar b}\right).
\ee
The infinitesimal $SU(n+1)/U(n)$
transformation of the coordinates is
\be\label{tr}
\delta z^a =
\varepsilon^a + \left(\bar \varepsilon\cdot z\right)  z^a  \, ,
\ee
where $\{\varepsilon^a; a=1,\dots,n\}$ are $n$ constant
complex parameters, and this induces the K\"ahler gauge
transformation
\be
\delta{\cal K} =  \left(\bar\varepsilon\cdot z
+ \varepsilon\cdot \bar  z \right).
\ee
This is manifestly an
infinitesimal isometry of the K\"ahler metric and a symmetry of
${\cal F}\,$. The $SU(n+1)/U(n)$ variation of the Lagrangian
(\ref{CPlag}) is a total  time derivative. The subgroup $U(n)\subset SU(n+1)$
is realized as linear transformations of $z^a, \bar z_a\,$.

With the standard notations for the conjugate  momenta, one finds that the classical Hamiltonian is
\be
H_{class}=g^{a {\bar b}}\left(p_a- N{\cal A}_a \right) \left(\bar p_{\bar b} - N{\cal A}_{\bar b} \right),
\ee
where the inverse metric is:
\be
g^{a \bar b} = \left(1 + \bar z\cdot z\right)\left[ \delta^{a \bar b} + z^a\bar z^b\right].
\ee
We quantize the model  via the standard replacement
\be
p_a\rightarrow -i \partial_a\, ,\qquad  \bar p_{\bar b} \rightarrow -i \partial_{\bar b}\, .
\ee
Defining the quantum Hamiltonian through symmetric ordering of the covariant derivatives one has
\be
H = -{1\over 2} g^{a {\bar b}} \left\{\nabla^{(N)}_{\bar b}, \nabla^{(N)}_a\right\} =
-g^{a {\bar b}}\nabla^{(N)}_a\nabla^{(N)}_{\bar b} + Nn\,,
\ee
where
\be
\nabla^{(N)}_a = \partial_a  -N\partial_a {\cal K} \, , \qquad
\nabla^{(N)}_{\bar b } = \partial _{\bar b}  + N\partial_{\bar b} {\cal K} \, .
\ee
These covariant derivatives have the commutation relation
\be
 \left[ \nabla^{(\tilde N)}_{\bar b}, \nabla^{(N)}_a\right] = -\left( N + \tilde N\right)g_{{\bar b} a}\, .
\ee

Now consider, for integer $2N \geq 0\,$,  the sequence of wave functions
\be\label{vector1pri}
\Psi^{(N)}_{\ell} = \nabla^{(N+n+1)}_{a_1} \nabla^{(N+n+3)}_{a_2} \cdots
\nabla^{(N +n+ 2\ell +1)}_{a_\ell} \Phi^{a_1 a_2 \cdots   a_{\ell}}\, ,
\ee
where $\Phi^{a_1 a_2 \cdots   a_{\ell}}$ is a totally symmetric
$\left(\ell,0\right)$ tensor satisfying the analyticity conditions
\be
\nabla^{(N)}_{\bar b}\Phi^{a_1 a_2 \cdots   a_{\ell}} = 0\, .
\ee
We claim that these are  eigenfunctions of $H$ with eigenvalue
\be
E_{\ell} = \ell\left(2N+n +\ell\right) +Nn\, .
\ee
The proof goes very much like the one outlined in Section \ref{sect:spectrum} of the text,
and rests on the identities
\bea
g^{a{\bar d}}\partial_c g_{{\bar d} b} &=& -\left[ 1+ \bar z\cdot z\right]^{-1}
\left[\delta_b^a {\bar z}^c +\delta_c^a{\bar z}^b\right], \nn
g^{d{\bar a}}\partial_{\bar c}g_{{\bar b} { d}} &=& -\left[ 1+ \bar z\cdot z\right]^{-1}
\left[\delta_b^a z^c +\delta_c^a z^b\right].
\eea
The total symmetry in the indices of $\Phi^{A_1 A_2 \cdots   A_{\ell}}\left(P\right)$
is necessary in order to obtain the simple form of the `semi-covariant'  derivatives
appearing in [\ref{vector1pri}], from the fully covariant ones. Of course, cohomology arguments
fix the value of $N$ to be a half integer but this can also be
deduced from convergence of the $SU(n+1)$-invariant norm
\bea
||\Psi||^2 = \int \prod_{a=1}^{n} dz^a d{\bar z}^a \, e^{-(n+1){\cal K}} \, | \Psi |^2\, .
\eea
One could pursue this analysis for $\bC\bb{P}^{\left(n|m\right)}$ but we will not attempt this here.

%%%%%%%%%%%%%%%%%%%%%%%%%%

\end{document}